\documentclass[a4paper,11pt]{article}
\pdfoutput=1 % if your are submitting a pdflatex (i.e. if you have
\usepackage{jheppub} % for details on the use of the package, please
\usepackage{amsmath,amssymb,amsthm,amscd,graphicx}
\usepackage[notref,notcite]{}

\input epsf.sty

\theoremstyle{definition}

\newcommand{\bs}{\boldsymbol}

\newcommand{\CC}{{\cal C}}
\newcommand{\CE}{{\cal E}}

\newcommand{\CI}{{\cal I}}
\newcommand{\CJ}{{\cal J}}

\newcommand{\CL}{{\cal L}}

\newcommand{\CO}{{\cal O}}

\newcommand{\CS}{{\cal S}}

\newcommand{\CV}{{\cal V}}

\def\IZ{{\mathbb Z}}

\newcommand{\mmm}{\mathsf{m}}

\newcommand{\re}{{\rm e}}
\newcommand{\ri}{{\rm i}}
\newcommand{\rd}{{\rm d}}

\newcommand{\be}{\begin{equation}}
\newcommand{\ee}{\end{equation}}
\newcommand{\ba}{\begin{aligned}}
\newcommand{\ea}{\end{aligned}}
\newcommand{\ben}{\begin{eqnarray}\displaystyle}
\newcommand{\een}{\end{eqnarray}}

\newcommand{\sectiono}[1]{\section{#1}\setcounter{equation}{0}}

\newdimen\tableauside\tableauside=1.0ex
\newdimen\tableaurule\tableaurule=0.4pt
\newdimen\tableaustep
\def\phantomhrule#1{\hbox{\vbox to0pt{\hrule height\tableaurule width#1\vss}}}
\def\phantomvrule#1{\vbox{\hbox to0pt{\vrule width\tableaurule height#1\hss}}}
\def\sqr{\vbox{%
  \phantomhrule\tableaustep
  \hbox{\phantomvrule\tableaustep\kern\tableaustep\phantomvrule\tableaustep}%
  \hbox{\vbox{\phantomhrule\tableauside}\kern-\tableaurule}}}
\def\squares#1{\hbox{\count0=#1\noindent\loop\sqr
  \advance\count0 by-1 \ifnum\count0>0\repeat}}
\def\tableau#1{\vcenter{\offinterlineskip
  \tableaustep=\tableauside\advance\tableaustep by-\tableaurule
  \kern\normallineskip\hbox
    {\kern\normallineskip\vbox
      {\gettableau#1 0 }%
     \kern\normallineskip\kern\tableaurule}%
  \kern\normallineskip\kern\tableaurule}}
\def\gettableau#1{\ifnum#1=0\let\next=\null\else
\squares{#1}\let\next=\gettableau\fi\next}

\def\bPhi{\boldsymbol{\Phi}}

\newcommand{\figref}[1]{Fig.~\protect\ref{#1}}
\title{\Huge{\boldmath A new renormalon in two dimensions}}

\author{Marcos Mari\~no and Tom\'as Reis}

\affiliation{D\'epartement de Physique Th\'eorique et Section de Math\'ematiques\\
Universit\'e de Gen\`eve, Gen\`eve, CH-1211 Switzerland}

\emailAdd{marcos.marino@unige.ch, tomas.reis@unige.ch} 

\abstract{According to standard lore, perturbative series of super-renormalizable theories have only instanton singularities. 
In this paper we show that two-dimensional 
scalar theories with a spontaneously broken $O(N)$ symmetry at the classical level, which are super-renormalizable, have an IR renormalon singularity at large $N$. 
Since perturbative expansions in these theories are made around the ``false vacuum" in which the global symmetry is broken, this singularity can be regarded 
as a manifestation of the non-perturbative absence of Goldstone bosons. We conjecture that the Borel singularity 
in the ground state energy of the Lieb--Liniger model is a non-relativistic manifestation of this phenomenon. We also 
provide {\it en passant} a detailed perturbative calculation of the Lieb--Liniger energy up to two-loops, and we check that it 
agrees with the prediction of the Bethe ansatz.}    

\begin{document}
\maketitle
\flushbottom
 
\sectiono{Introduction}

The study of the large order behavior of perturbative series in quantum theory has provided an efficient window 
on non-perturbative phenomena. In quantum mechanics, it was found in \cite{lam, bw2} and in many 
subsequent works that this behavior is controlled by instantons, and is due to the factorial 
growth in the number of Feynman diagrams \cite{bw-stat} 
(see e.g. \cite{mmbook} for a textbook introduction). In quantum field theory, the situation is more complicated, since in many theories 
one can find specific diagrams which grow factorially 
with the loop order after integration over the momenta \cite{gross-neveu,lautrup,parisi1,parisi2, thooft}. These diagrams are 
usually called {\it renormalon diagrams} (see \cite{beneke} for an extensive review). 
They lead to singularities in the Borel plane of the coupling constant which, following \cite{beneke}, we will call 
renormalon singularities, or renormalons for short. Depending on the region 
in momenta which leads to the factorial growth, one has UV or IR renormalons. In asymptotically free theories and in QED,  
renormalons are believed to control the large order behavior of perturbation theory. Evidence for this was found in \cite{pineda, pineda2} in the 
case of Yang--Mills theory, and in \cite{fkw,volin, pcf-lattice,mr-ren} for integrable two-dimensional theories. 

It is often stated in the literature that renormalons, as their name indicate, are typical of renormalizable field theories, 
while super-renormalizable field theories only have instanton singularities (see e.g. \cite{parisi1,lgzj,zjbook}). A typical example of the latter case is the two-dimensional 
field theory for a real scalar field $\Phi$ with Lagrangian 
\be
\label{z2-version}
\CL(\Phi)= {1\over 2} \partial_\mu \Phi\partial^\mu \Phi- {\mu^2 \over 2} \Phi^2 -{g\over 4!} \Phi^4. 
\ee
This theory is super-renormalizable. Its perturbative expansion in powers of the dimensionless coupling $g/m^2$ 
is factorially divergent but Borel-summable \cite{eckmann, blgzj-ins,serone1, serone2}. Its large order behavior is controlled by 
instanton singularities, both in the phase where $\mu^2>0$ \cite{brezin-parisi,serone1}, as well as in the phase where 
$\mu^2<0$ and the $\IZ_2$ symmetry is spontaneously broken \cite{serone2}.

We can promote the field $\Phi$ to an $N$-dimensional vector $\bPhi=(\Phi_1, \cdots, \Phi_N)$ and consider the 
version of (\ref{z2-version}) with a global 
non-Abelian symmetry $O(N)$. The theory is still super-renormalizable, but the physics when $\mu^2<0$ is very different: a 
famous theorem by Coleman, Mermin and Wagner \cite{coleman, mw} 
states that there are no Goldstone bosons in two dimensions, therefore the $O(N)$ 
symmetry can not be broken quantum-mechanically. One symptom of this situation is that, when $\mu^2<0$, 
the quantum corrections to the vacuum expectation value (vev) of $\bPhi$ are afflicted with IR divergences. In a sense, 
the classical vacuum in which $\langle \bPhi  \rangle \not=0$ can be regarded as a ``false vacuum" in 
the full quantum theory. In spite of this, it was shown by Jevicki 
\cite{jevicki} that, after renormalization and an appropriate IR regularization, 
the perturbative expansion for the ground state energy is well-defined, and IR 
divergences cancel order by order in the expansion in the coupling constant. This 
result was extended in \cite{elitzur, david} to all $O(N)$-invariant correlation functions. 

In this paper we argue that the perturbative series for the ground state energy of the $O(N)$ theory 
around this false vacuum is factorially divergent and leads to a Borel singularity in the positive real axis. Therefore, the series 
is not Borel summable, and the Borel singularity turns out to be an IR renormalon. This is then an 
example of an IR renormalon singularity in a super-renormalizable theory. 

In order to establish the existence of a renormalon in this theory, we consider the leading contribution 
to the perturbative series at large $N$, which is due to a family of bubble-like diagrams. We show that, in agreement with Jevicki's result, 
their contribution is finite, order by order, but it grows factorially due to the momentum integration in the IR region. 
Our strategy is very similar to the analysis of renormalons in QED and QCD. In these theories, one can take a large $N_f$ or large $\beta_0$ 
limit, respectively, in which perturbation theory is dominated by bubble chains with a renormalon behavior \cite{beneke}. 
The renormalon singularity we find at large $N$ is very simple and it can be analysed in 
detail with the tools of the theory of resurgence \cite{mmbook, mmlargen, abs-primer}.  
It is in fact possible to calculate the exact Borel transform of the divergent series associated to the renormalon diagrams,  
and to determine the full trans-series incorporating non-perturbative corrections.

The motivation for this work came from the the study of a non-relativistic analogue of the two-dimensional $O(N)$ scalar theory, 
namely, the Lieb--Liniger model \cite{ll}. 
This model describes a Bose gas in one spatial dimension with a delta function repulsion, and the ground state energy can 
be found exactly with the Bethe ansatz. In \cite{mr-ll} we obtained analytic results for the perturbative series of the ground state energy up to very 
large order and we observed that it diverges factorially, leading to a Borel singularity in the positive real axis. 
Instanton solutions leading to this singularity do not seem to exist. However, as we will see in this paper, the 
perturbative expansion of the Lieb--Liniger model is very similar to the one in the $O(N)$ model (this was already pointed out in \cite{jevicki}). 
It is then natural to conjecture that the Borel singularity found in the Lieb--Liniger model is an IR renormalon, 
similar to the one found in the relativistic $O(N)$ theory. We have not been able to provide a concrete realization of this scenario by finding an appropriate family of renormalon diagrams, but we go 
through the exercise of computing the ground state energy at two-loops by using the 
field theory approach of \cite{b-nieto}. We show that the IR divergences in the Lieb--Liniger model have 
the same structure than the ones in the relativistic $O(N)$ theory, and we verify that they cancel up to two-loops. 
As an aside, we check that the final result agrees with the answer obtained from the Bethe ansatz\footnote{The agreement with the Bethe ansatz up to two-loops 
was verified long ago in the 
very different hydrodynamic formalism of \cite{popov,taka-nnl}, which does not have IR divergences.}. 

This paper is organized as follows. In section \ref{sec2}, after reviewing the results of \cite{jevicki}, 
we present our main calculation, namely, the ground state energy of the $O(N)$ 
model at next-to-leading order in the $1/N$ expansion. We show that the resulting expansion in the 
coupling constant leads to a factorially divergent series and an IR renormalon, and we perform a detailed resurgent 
analysis of the series. 
We also generalize the result to any potential for the scalar field. In section \ref{sec-ll} we study the 
Lieb--Liniger model up to two-loops by using field theory techniques. 
Finally, in \ref{sec-con} we conclude and list some problems for the future.

\sectiono{Renormalons in the two-dimensional $O(N)$ model}

\label{sec2}
\subsection{The ground state energy at two-loops}
\label{sec-2loop}

In this paper we will focus on the perturbative series for the ground state energy of the two-dimensional $O(N)$ model. A good starting 
point for our analysis is a review of the two-loop calculation due to Jevicki \cite{jevicki} in the case of a quartic potential, since 
our main calculation is a large $N$ generalization of Jevicki's result. In these calculations, 
one performs an expansion around the classical vacuum in which the symmetry is spontaneously broken. 
As a consequence of the Coleman--Mermin--Wagner theorem, 
this is a ``false vacuum" in the full quantum theory. Jevicki showed that, although 
IR divergences appear in intermediate steps of the calculation, they cancel in the final answer. 

Let us then consider the standard $O(N)$ theory for a vector field $\bPhi=(\Phi_1, \cdots, \Phi_N)$, described by the Lagrangian
\be
\label{lag}
\CL(\bPhi)= {1\over 2} \partial_\mu \bs{\Phi}\cdot \partial^\mu \bs{\Phi}-V(\bPhi), 
\ee
where the potential is given by
\be
V(\bs{\Phi})= {\mu^2 \over 2} \bs{\Phi}^2 +{g\over 4!} \bs{\Phi}^4. 
\ee
We denote $\bPhi^4= (\bPhi^2)^2$. We will consider the stable case in which $g>0$. Classically, there are two phases. 
When $\mu^2>0$, the absolute minimum occurs at $\bs{\Phi}=0$. This is the symmetric phase. The phase of spontaneously broken symmetry corresponds to 
\be
\mu^2 <0. 
\ee
In this case, the minimum occurs at 
\be
\label{classm}
\bs{\Phi}^2 =\phi_\star^2= - 3! {\mu^2 \over g}. 
\ee
Classically, the $O(N)$ symmetry is spontaneously broken down to $O(N-1)$. 

\begin{figure}[!ht]
\leavevmode
\begin{center}
\includegraphics[height=1cm]{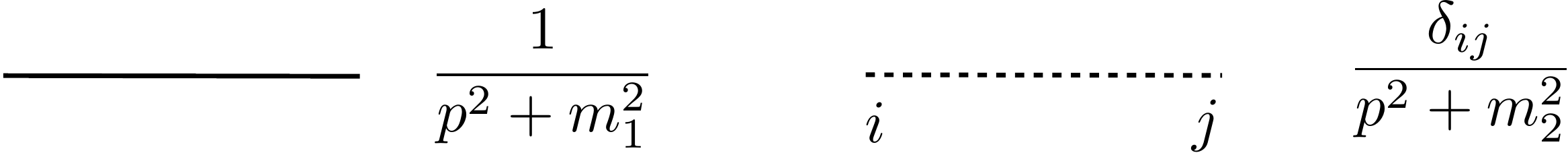}
\end{center}
\caption{The propagators for the $\xi$ and the $\bs{\eta}$ fields.}
\label{f-propas}
\end{figure}

Quantum mechanically, to determine the ground state and the ground state energy, we have to compute the Coleman--Weinberg effective potential \cite{cw}. We will follow
\cite{jackiw-f}: first we split $\bs{\Phi}$ as
\be
\label{shift}
\bs{\Phi}= \left( \phi, 0, \cdots, 0 \right) + (\xi, 0, \cdots, 0)+ (0, \eta_1, \cdots, \eta_{N-1}), 
\ee
where $\phi$ is a constant field configuration, while $\xi$, $\bs{\eta}=(\eta_1, \cdots, \eta_{N-1})$ are quantum fluctuations. 
We will calculate $V(\phi)$ as the sum of all one-particle irreducible diagrams. The quantum corrected vev $\phi$ is then determined by 
\be
\label{min-effpot}
{\partial V \over \partial \phi}=0, 
\ee
and the ground state energy is obtained by evaluating the effective potential on the solution of (\ref{min-effpot}). 
From now on we will work in the Euclidean theory. The relevant Lagrangian is 
\be
\label{final-lag-E}
\ba
\CL (\phi, \xi, \bs{\eta})&={1\over 2} \mu^2 \phi^2+{g \over 4!} \phi^4 \\
&+ {1\over 2} \partial_\mu\xi  \partial^\mu \xi +{m_1^2 \over 2} \xi^2 + 
 {1\over 2} \partial_\mu \bs{\eta}\cdot \partial^\mu \bs{\eta}+{m_2^2 \over 2} \bs{\eta}^2 \\
 &+{g\over 3!} \phi  \xi^3 +{g \over 3!} \phi \xi \bs{\eta}^2 + {g \over 12} \xi^2 \bs{\eta}^2 +{g \over 4!} \xi^4 +{g \over 4!} \bs{\eta}^4.
 \ea
 \ee
 We have removed the linear vertex in $\xi$, following the prescription of \cite{jackiw-f}. In the second line of (\ref{final-lag-E}) we have introduced the masses
\be
\label{regul1}
\ba
m_1^2 &= \mu^2 +{g\over 2} \phi^2, \qquad m_2^2 &= \mu^2 + {g \phi^2 \over 3!} + \epsilon^2. 
\ea
\ee
Following Jevicki \cite{jevicki}, we have introduced an IR regulator $\epsilon^2$ which will be taken to zero at the end of the calculations. 
When evaluated at the classical minimum (\ref{classm}), we have 
\be
\label{regul2}
m_1^2 =-2 \mu^2 := m^2, \qquad m_2^2 =\epsilon^2. 
\ee
In the limit $\epsilon \rightarrow 0$, the $N-1$ scalar fields $\eta_i$, $i=1, \cdots, N-1$ become massless 
and are the classical Goldstone bosons of the model. 
The propagator of the $\xi$ field is represented by a full line, while the one of the $\bs{\eta}$ fields is represented a dashed line; they are shown in \figref{f-propas}. 

\begin{figure}[!ht]
\leavevmode
\begin{center}
\includegraphics[height=3cm]{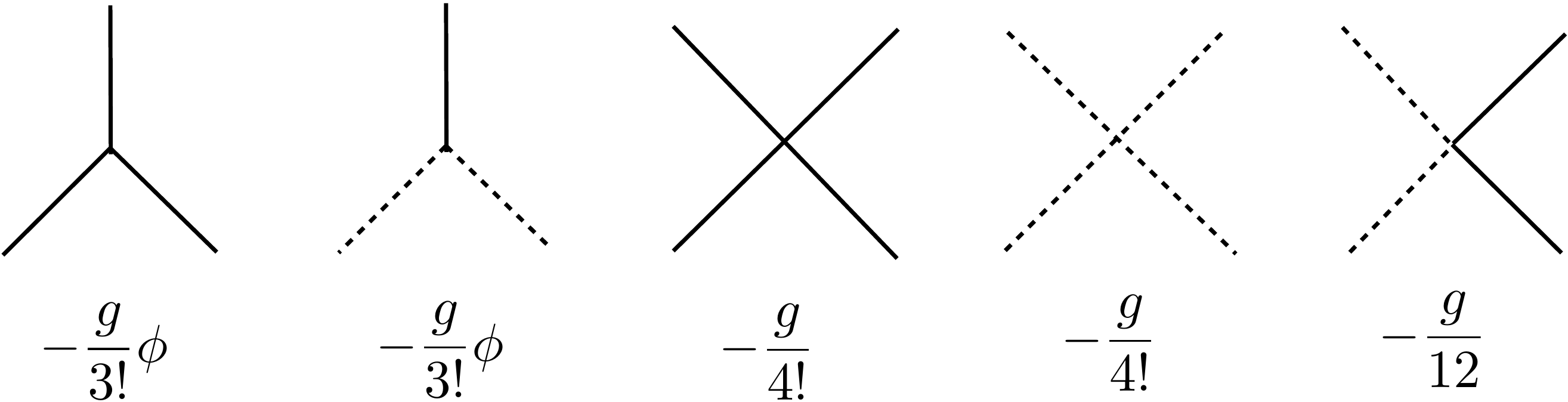}
\end{center}
\caption{The vertices for the spontaneously broken phase of the $O(N)$ model.}
\label{f-vertices}
\end{figure}

Finally, the third line in (\ref{final-lag-E}) gives the interaction terms. This leads to five types of Feynman vertices which we represent in \figref{f-vertices}. 

In order to keep track of the loop order we introduce a $\hbar$ parameter so that the effective potential reads, up to two-loops, 
\be
V(\phi^2)= {1\over \hbar} V_0(\phi^2)+ V_1(\phi^2)+\hbar  V_2(\phi^2)+ \cdots
\ee
The tree-level and one-loop contributions are given by
\be
\label{tone}
\ba
V_0 (\phi)&= {1\over 2} \mu^2 \phi^2+{g \over 4!} \phi^4, \\
V_1(\phi) &={1\over 2}  \int {\rd^d k \over (2 \pi)^d} \log(k^2+ m_1^2) +{1\over 2} (N-1) \int {\rd^d k \over (2 \pi)^d} \log(k^2+ m_2^2). 
\ea
\ee
 \begin{figure}[!ht]
\leavevmode
\begin{center}
\includegraphics[height=4cm]{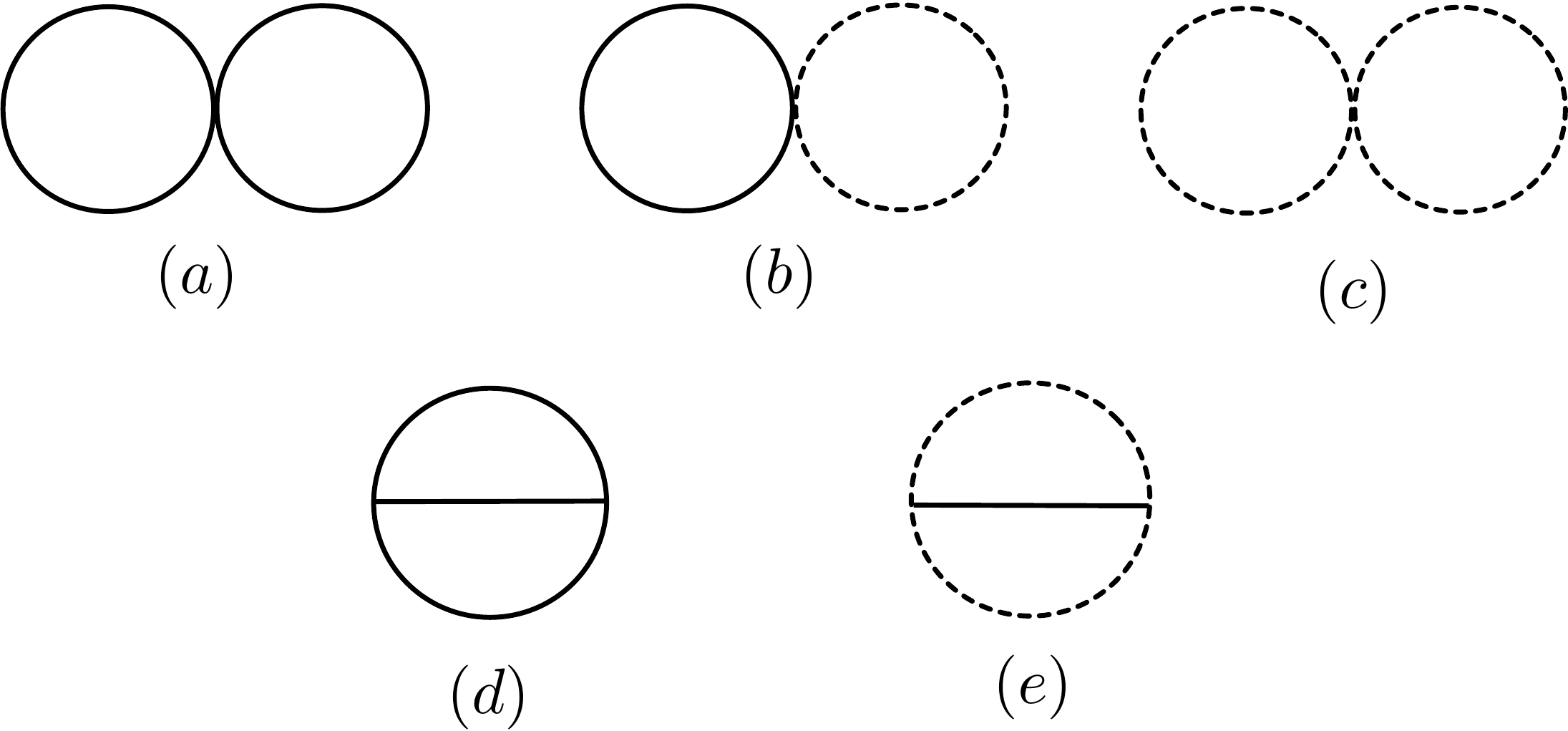}
\end{center}
\caption{Diagrams contributing to the two-loop effective potential .}
\label{f-2loops}
\end{figure} 
The two-loop contribution is given by the sum of the diagrams in \figref{f-2loops}. We will write it as
\be
V_2(\phi)=V_2^{(1)}(\phi)+ V_2^{(2)}(\phi), 
\ee
where $V_2^{(1)}$ only involves products of one-loop integrals, while $V_2^{(2)}$ involves genuine two-loop integrals. We have, 
\be
V_2^{(1)}= (a)+ (b) +(c), 
\ee
and
\be
\ba
(a)&=3  {g\over 4!} \left( \int {\rd^d k \over (2 \pi)^d} {1\over k^2+ m_1^2} \right)^2, \\
(b)&=2 (N-1) {g\over 4!} \left( \int {\rd^d k \over (2 \pi)^d} {1\over k^2+ m_1^2} \right)\left( \int {\rd^d k \over (2 \pi)^d} {1\over k^2+ m_2^2} \right),\\
(c)&= (N^2-1) {g\over 4!}\left( \int {\rd^d k \over (2 \pi)^d} {1\over k^2+ m_2^2} \right)^2. 
\ea
\ee
On the other hand, 
\be
V_2^{(2)}=(d)+ (e), 
\ee
where 
\be
\ba
(d)&=-{3! \over 2} \left( {g \over 3!} \right)^2 \phi^2  \int {\rd^d k \over (2 \pi)^d} {\rd^d l \over (2 \pi)^d} {1\over k^2+ m_1^2}{1\over l^2 + m_1^2}{1\over (k+l)^2+ m_1^2}, \\
(e)&= - \left( {g \over 3!} \right)^2 \phi^2 (N-1)  \int {\rd^d k \over (2 \pi)^d} {\rd^d l \over (2 \pi)^d} {1\over k^2+ m_2^2}{1\over l^2 + m_2^2}{1\over (k+l)^2+ m_1^2}. 
\ea
\ee
We now solve the minimization equation (\ref{min-effpot}) order by order in the loop expansion, as
\be
\phi^2= \phi^2_\star + \hbar \phi^2_1+ \hbar^2 \phi^2_2+ \cdots,
\ee
where $\phi_\star^2$ is the classical value (\ref{classm}). The first quantum correction is 
\be
\label{phi1}
\phi^2_{1}= -\left( {\partial^2 V_0 \over \partial \left(\phi^2\right)^2}(\phi^2_\star) \right)^{-1} {\partial V_1 \over \partial \phi^2} (\phi^2_\star). 
\ee
The final result for the ground state energy at two-loops is 
\be
\label{e-2loop}
\ba
E(m^2, g)&= {1\over \hbar} V_0(\phi^2_\star)+ V_1(\phi^2_\star) + \hbar \left( V_2(\phi^2_\star)- {1\over 2} {\partial^2 V_0 \over \partial \left(\phi^2\right)^2} (\phi^2_\star) 
\phi^4_1 \right)+ \cdots
\ea
\ee

So far we have not evaluated the Feynman integrals, which display both UV divergences as well 
as IR divergences when the IR regulator $\epsilon$ is taken to zero. 
UV divergences in scalar field theories in two dimensions can be renormalized by normal-ordering the fields. 
In the case of an $N$-dimensional vector, one has
\be
\ba
:\bPhi^2:&= \bPhi^2- N \CI(\mmm), \\
:\bPhi^4:&= \bPhi^4- 2(N+2) \CI(\mmm)\bPhi^2+N(N+2) \CI^2(\mmm),
\ea
\ee
where
\be
\CI(\mmm) = \int  {\rd^d k \over (2 \pi)^d} {1\over k^2+ \mmm^2}, 
\ee
and $\mmm$ parametrizes the choice of mass in the normal ordering. Different choices of $\mmm$ correspond to different 
renormalization schemes. We note that $\mmm$ is not necessarily the mass of the scalar field appearing in the theory. Normal-ordering renormalization is 
equivalent to renormalizing the mass and the vacuum energy in the original Lagrangian (\ref{lag}) 
(see the recent discussions in \cite{slava,serone1, serone2, serone-scheme}). The mass renormalization is 
\be
\label{mass-renor}
\mu^2=\tilde \mu^2- g {N+2 \over 6} \CI(\mmm), 
\ee
where $\mu$ is the bare mass and $\tilde \mu$ is the renormalized mass. In addition, one has to add to the vacuum energy the counterterm 
\be
g {N(N+2)  \over 4!} \CI^2(\mmm). 
\ee
One property of normal ordering is that, if there is an interacting scalar field of mass $\mmm$, diagrams involving contractions of legs in the same vertex 
vanish. Therefore, a convenient choice, made in \cite{jevicki}, is to do normal ordering w.r.t. the mass of the $\xi$ field in the classical vacuum, 
i.e. to choose
\be
\mmm^2=m^2. 
\ee
This is the scheme also used in the recent works \cite{serone1,serone2}. 
In practice, this means that the UV divergent one-loop integrals appearing in the calculations above are renormalized as
\be
\label{reg-int}
\int  {\rd^2 k \over (2 \pi)^2}{1\over k^2+a^2}\rightarrow \int  {\rd^2 k \over (2 \pi)^2}\left( {1\over k^2+a^2}-{1\over k^2+m^2} \right)=-{1\over 4 \pi} \log \left( {a^2 \over m^2} \right). 
\ee
With this choice of renormalization scheme, the diagrams (a) and (b) vanish. 
We renormalize the one-loop effective potential (i.e. the vacuum energy) in a way which is consistent with (\ref{reg-int}), namely 
  \be
  \label{log-int}
  \int {\rd^2 k \over (2 \pi)^2} \log(k^2+ a^2) \rightarrow    { a^2 \over 4 \pi}  \left(1- \log \left( {a^2 \over m^2} \right) \right).
  \ee
In particular, (\ref{tone}) reads,
\be
V_1 (\phi_\star^2) = { m^2 \over 8 \pi} + {N-1\over 8\pi}  \epsilon^2  \left(1- \log \left( {\epsilon^2 \over m^2} \right) \right).
\ee
All UV divergences are removed by this procedure. However, we still have to take care of possible IR divergences 
as we send the IR cutoff to zero: $\epsilon^2\rightarrow 0$. 
Some of the quantities we have computed are genuinely IR divergent, like for example the one-loop correction to the vev, 
\be
\label{phi1loop}
\phi^2_1= {N-1 \over 4 \pi} \log \left({\epsilon^2 \over m^2} \right). 
\ee
This is a reflection of the Coleman--Mermin--Wagner theorem, 
namely the classical vev is destabilized by 
IR divergences in the quantum corrections. 
However, Jevicki observed that when the IR regulated results are 
plugged into (\ref{e-2loop}), 
the IR divergences {\it cancel} and the limit $\epsilon^2 \rightarrow 0$ 
leads to a finite result. To see this, we note that the Feynman integral appearing in 
the diagram (e) is given by 
\be
\ba
& \int {\rd^2 k \over (2 \pi)^2} {\rd^2 l  \over (2 \pi)^2} {1\over k^2+ \epsilon^2} {1\over l^2+ \epsilon^2} {1\over (k+l)^2+ m^2} = {1\over (4 \pi m)^2} \left( 
\left(\log\left({\epsilon^2 \over m^2} \right)\right)^2  + {\pi^2 \over 3} \right)+ \CO(\epsilon^2).
\ea
\ee
It is now easy to see that the IR divergences of the form $\log^2(\epsilon^2)$ 
appearing in $V_2(\phi_\star^2)$ cancel against the term involving $\phi_1^4$ in the ground-state energy. 
Finally, the integral appearing in (d) is finite and given by  (see e.g. \cite{two-loop})
\be
\int {\rd^2 k \over (2 \pi)^2} {\rd^2 l  \over (2 \pi)^2} {1\over k^2+ m^2} {1\over l^2+ m^2} {1\over (k+l)^2+ m^2} ={1\over (4 \pi)^2} {2 A\over m^2} , 
\ee
where
\be
\label{jev-constant}
A ={\sqrt{3}} \, {\rm Im}\, {\rm Li}_2\left( \re^{2 \pi \ri/3} \right). 
\ee
Putting all together, we finally obtain Jevicki's result \cite{jevicki}
\be
\label{e2loops}
\CE (\widehat g, N)= -{3 \over 8 \widehat g} + {1\over 8\pi}- \left({A \over 2 }+{N-1\over 12} {\pi^2 \over 3}  \right){ \widehat g \over  (4 \pi)^2} 
+ \CO(
\widehat g^2), 
\ee
where we have used the dimensionless quantities 
\be
\label{hat-g}
\widehat g={g \over m^2}, \qquad \CE ={E \over m^2}. 
\ee
 The well-defined series (\ref{e2loops}) is expected to give the correct asymptotic 
expansion for the ground state energy of the theory (\ref{lag}) with $\mu^2<0$. 
As already noted by Jevicki in \cite{jevicki}, this is confirmed by a non-relativistic analogue of this theory with an 
$O(2)$ global symmetry: the Lieb--Liniger model \cite{ll}. In that model, one calculates the perturbative series for the 
ground state energy by expanding around a classical vacuum which breaks the $O(2)$ symmetry. The resulting series 
is not only well-defined, but it gives the correct asymptotic expansion of the 
ground state energy, as it can be verified directly by comparing it to the weak-coupling expansion of the exact solution 
obtained with the Bethe ansatz. We will perform this detailed check and comparison up to two-loops in our study of the Lieb--Liniger model in section \ref{sec-ll}.

\subsection{Large $N$ expansion and renormalons}

The ground state energy density, which is given at two-loops in (\ref{e2loops}), depends on $N$ and $\hat g$. By using 
large $N$ counting, it is easy to show that its perturbative expansion in powers of $\hat g$ has the following structure,
\be
\label{erk1}
\CE(\widehat g, N)= -{3 \over 8 \widehat g}+ \sum_{k \ge 0} \left( \sum_{r=1}^{1+k} e_{r, k} N^{1+k-r} \right) \widehat g^k.  
\ee
In particular, each coefficient in this series is a polynomial in $N$. The calculation of the full perturbative 
series (\ref{erk1}) to high loop order is difficult, but one can consider the limit in which $N$ is large. In this case, 
at each order in $\hat g$, the leading contribution comes from the coefficients $e_{1,k}$. As we 
will see, this sequence of coefficients is associated 
to a very specific type of diagrams and it can be 
calculated in closed form. It grows factorially with $k$, and leads to a Borel singularity in the positive real axis. This implies the 
existence of a renormalon singularity in the original perturbative series, at least if $N$ is large enough. This strategy to 
study the renormalon structure of $\CE(\widehat g, N)$ is similar to the large $N_f$ or large $\beta_0$ limit of the perturbative 
expansion in QED and QCD, respectively. In this limit, perturbation theory is dominated by bubble diagrams, and this can be used to 
establish the existence of renormalons in these theories \cite{beneke}. Renormalons 
in the scalar $O(N)$ theory in four dimensions have been also studied by calculating the effective potential at 
next-to-leading order in the large $N$ expansion \cite{galli}.

In order to further understand the structure of (\ref{erk1}) at large $N$, it is convenient to introduce 
the 't Hooft coupling
\be
\lambda= g N
\ee
and its dimensionless counterpart, 
\be
\widehat \lambda= \widehat g N, 
\ee
and reorganize the expansion as 
\be
\label{enln}
\CE(\widehat g, N)=\sum_{r\ge 0}N^{1-r} \CE_{(r)}(\widehat \lambda), 
\ee
where
\be
\label{leadinge0}
\CE_{(0)}(\widehat \lambda)=-{3 \over 8 \widehat \lambda}
\ee
and
\be
\label{erk2}
 \CE_{(r)}(\widehat \lambda)=  \sum_{k \ge 0} e_{r, k} \widehat \lambda^k. 
 \ee
 In the large $N$ limit at fixed 't Hooft coupling, the first non-trivial contribution to the energy is 
 given by $\CE_{(1)}$, which encodes the coefficients $e_{1,k}$ for all $k$. From the point of view of the 
 $1/N$ expansion, this is the next-to-leading contribution to the energy density, and it can be calculated 
 from the $1/N$ expansion of the effective potential. 
 
The leading large $N$ contribution to the effective potential in dimensions $1\le d\le 4$ was obtained  in the well-known paper \cite{cjp}, and the subleading corrections were discussed in \cite{root}. Let us then review the relevant large $N$ techniques developed in \cite{cjp,root}. The first step is to 
perform a Hubbard--Stratonovich transformation. This leads to an 
 equivalent theory with two fields, $\bPhi$ and $X$, and Lagrangian, 
\be
\ba
\label{lag-chi}
\CL (\bPhi, X)&= \CL(\bPhi) +{3N \over 2 \lambda} \left(X- {\lambda \over 6 N} \bPhi^2 -\mu^2\right)^2\\
&= {1\over 2} \partial_\mu \bPhi \cdot \partial^\mu \bPhi -{X \over 2} \bPhi^2 + {3 N \over 2 \lambda} X^2 -{3N \mu^2 \over \lambda} X + {3N  \mu^4 \over 2 \lambda}. 
\ea
\ee
This theory is equivalent to the original one since $X$ is an auxiliary field which can be integrated out.   
To obtain the effective potential, we expand around a constant configuration of the fields, 
\be
\Phi_1(x) = \xi (x)+{\sqrt{N}} \phi, \qquad X(x)= \chi + { \tilde \chi (x) \over {\sqrt{N}}}, \qquad \Phi_j(x)= \eta_{j-1}(x), \qquad j=2, \cdots, N. 
\ee
After removing the linear terms and going to Euclidean signature, 
we obtain the Lagrangian 
\be
\label{shifted-lag-e}
\ba
\CL (\bs{\eta},  \xi, \tilde \chi; \phi, \chi)&= V_{\rm tree} (\phi, \chi) +  {1\over 2}  \partial_\mu \bs{\eta} \cdot \partial^\mu \bs{\eta} +{1\over 2} \chi  \bs{\eta}^2 \\ 
&+{1\over 2} \partial_\mu \xi \cdot \partial^\mu \xi +{1\over 2} \chi \xi^2 - {3 \tilde \chi^2 \over2 \lambda} + \tilde \chi \xi \phi  \\
&+{1\over 2{\sqrt{N}}} \tilde \chi \xi^2 +{1\over 2{\sqrt{N}}}  \tilde \chi \bs{\eta}^2.
\ea
\ee
In this expression, 
\be
V_{\rm tree} (\phi, \chi)= N \left( -{3 \chi^2 \over 2 \lambda} +{\chi \phi^2 \over 2} +{3 \mu^2 \chi \over \lambda} -{3 \mu^4 \over 2 \lambda} \right) 
\ee
is the tree level contribution. The effective potential has the large $N$ expansion
\be
V(\phi, \chi) = N \sum_{k \ge0} V_{(k)} (\phi, \chi) N^{-k}, 
\ee
and the minimization conditions 
\be
\label{min-two}
{\partial V \over \partial \phi}= {\partial V \over \partial \chi}=0 
\ee
can be also solved order by order in the $1/N$ expansion, 
\be
\phi= \phi_{(0)}+ {1\over N} \phi_{(1)}+ \cdots, \qquad \chi= \chi_{(0)}+ {1\over N} \chi_{(1)}+ \cdots
\ee
At large $N$, the effective potential is given by the tree level contribution, plus the one-loop contribution of the $N-1$ $\eta$ fields. 
We find in this way, after using the renormalization scheme discussed in section \ref{sec-2loop}, 
\be
\label{leadingV}
V_{(0)}(\phi, \chi)=  -{3 \chi^2 \over 2 \lambda} +{\chi \phi^2 \over 2} +{3 \mu^2 \chi \over \lambda} -{3 \mu^4 \over 2 \lambda} +{1 \over 8 \pi} \chi \left(1-\log\left( {\chi \over m^2} \right)\right),
\ee
which is the result of Coleman, Jackiw and Politzer \cite{cjp}\footnote{The renormalization of the mass (\ref{mass-renor}) due to normal ordering coincides at large $N$ with the 
renormalization scheme used in \cite{cjp}.}. 

In order to calculate the coefficients $e_{1,k}$ in (\ref{erk1}), we have to expand the effective potential around the saddle point which corresponds to the conventional perturbative vacuum. The minimization conditions (\ref{min-two}) give  
\be
\label{leading-phi}
\phi_{(0)}^2= { 6 \chi_{(0)} \over \lambda} -{6 \mu^2 \over \lambda} +{1\over 4 \pi} \log\left( {\chi_{(0)} \over m^2} \right), 
\ee
as well as 
\be
\chi_{(0)}=0. 
\ee
This leads to an IR divergence when plugged in (\ref{leading-phi}), which was interpreted in \cite{cjp} as 
a manifestation of the Coleman--Mermin--Wagner theorem. However, we can regulate this IR divergence as in \cite{jevicki}, by setting
\be
\label{leading-chi}
\chi_{(0)}= \epsilon^2, 
\ee
and taking the limit $\epsilon \rightarrow 0$ at the end of the calculation. By plugging this in (\ref{leading-phi}), we find the large 
$N$ limit of the result (\ref{phi1loop}) obtained in the loop expansion. 
Therefore, the saddle-point which makes contact with the conventional, IR-regularized perturbative 
expansion, is given by (\ref{leading-phi}), (\ref{leading-chi})\footnote{One could argue that the expansion should be done 
around the ``true" vacuum at large $N$, which is located at $\phi=0$. However, such an expansion would 
not be connected to the perturbative expansion (\ref{erk1}) around the classical vacuum, which is what we want to study here. This is similar to what happens in Fermi gases with an 
attractive interaction. There, one can expand around a non-trivial large $N$ vacuum with a gap, which is useful to study the theory at strong coupling, or one can expand around the perturbative vacuum, which leads to a sum over ladder diagrams \cite{nsr} with renormalon behavior \cite{mr-long}. We would like to thank Marco Serone for illuminating conversations on this point.}.

 \begin{figure}[!ht]
\leavevmode
\begin{center}
\includegraphics[height=2.5cm]{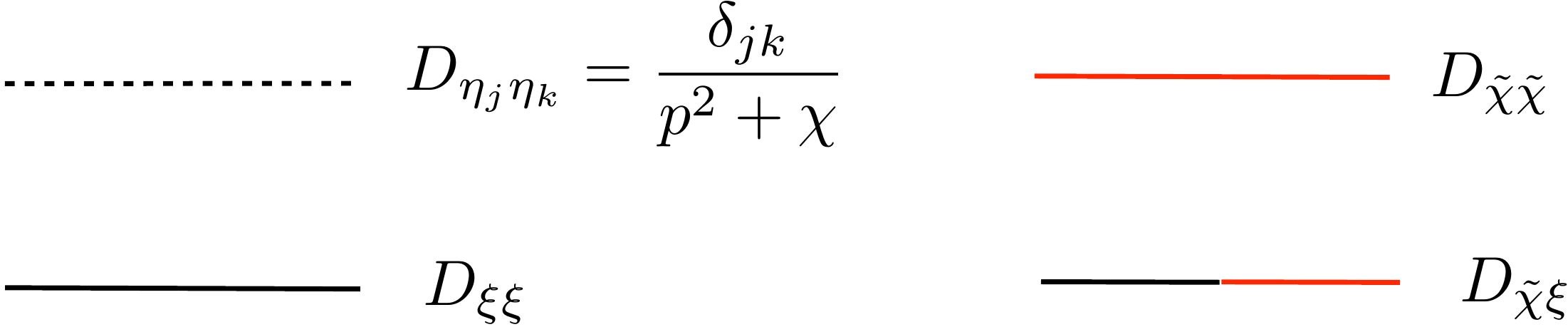}
\end{center}
\caption{The propagators for the effective Lagrangian (\ref{shifted-lag-e}).}
\label{f-eprop}
\end{figure}

\begin{figure}[!ht]
\leavevmode
\begin{center}
\includegraphics[height=3.5cm]{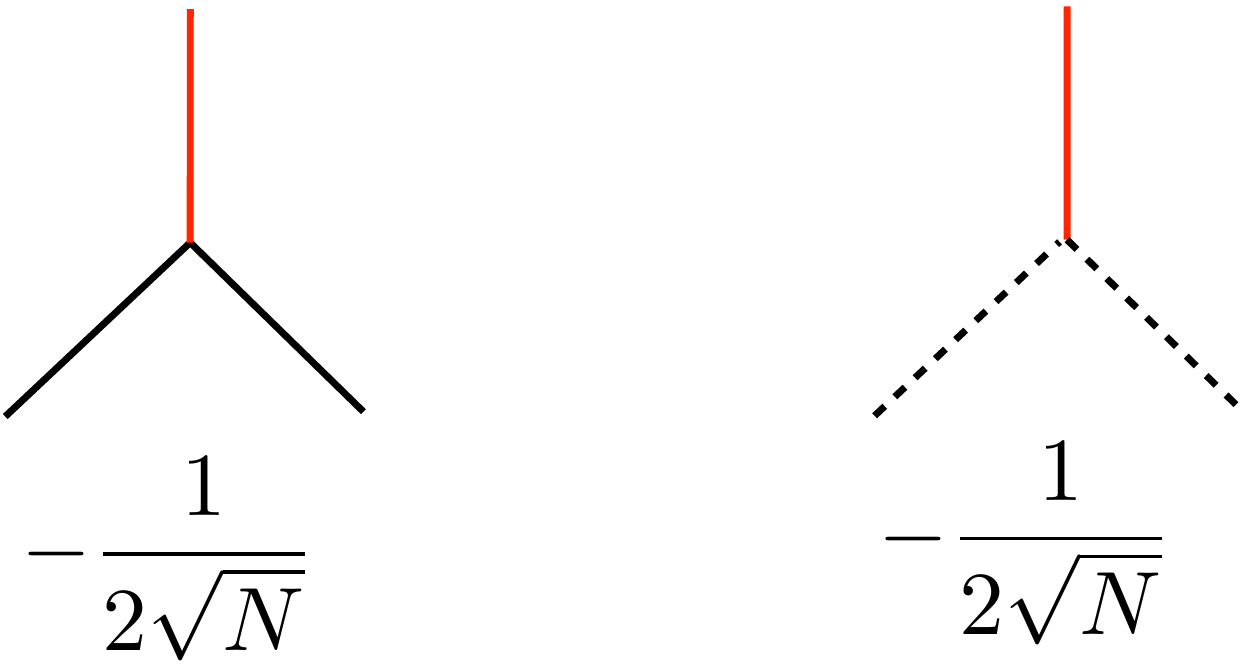}
\end{center}
\caption{The vertices for the effective Lagrangian (\ref{shifted-lag-e}). The first one represents the $\tilde \chi \xi^2$ coupling, while the second 
one represents the $\tilde \chi \eta^2$ coupling.}
\label{f-evertex}
\end{figure}

Let us now calculate the next-to-leading correction in $1/N$ to the effective potential, folllowing \cite{root}. 
We can think about (\ref{shifted-lag-e}) as a theory of the fields $\bs{\eta}$, $\tilde \chi$ and $\xi$, with a non-diagonal 
propagator for $\xi$, $\tilde\chi$. The inverse propagator for $\xi$, $\tilde \chi$ is given by the matrix
\be
\begin{pmatrix} p^2 + \chi & \phi \\ \phi & -3 /\lambda \end{pmatrix}
\ee
After inversion, we find the following propagators:
\be
\ba
D_{\xi \xi}&= {1\over p^2 + \chi +{\lambda \phi^2 / 3}}, \\
D_{\xi \tilde \chi}&= {\lambda \phi \over 3} {1 \over  p^2 + \chi +{\lambda \phi^2 / 3}}, \\
D_{\tilde\chi \tilde\chi}&= -{\lambda \over 3} {p^2 + \chi \over  p^2 + \chi +{\lambda \phi^2 / 3}}. 
\ea
\ee
In addition, the propagator for $\eta_j\eta_k$ is given by 
\be
D_{\eta_j\eta_k}= {\delta_{jk} \over p^2+ \chi}. 
\ee
These propagators are represented graphically in \figref{f-eprop}. The $\xi$, $\eta$, $\tilde \chi$ fields are represented by a full black line, dashed black line and full red line, 
respectively. In addition, we have the cubic vertices shown in \figref{f-evertex}. 

At one-loop, the $\eta$ fields gave a contribution of order $N$ in (\ref{leadingV}), but since there are only $N-1$ of them they also give a subleading contribution in the $1/N$ expansion, 
\be
- {1\over 2} \int{\rd^d k \over (2 \pi)^d} \log (k^2 + \chi). 
 \ee
The fields $\xi$, $\tilde\chi$ are coupled and they give the following contribution of order one to the effective potential:
\be
\label{half}
 {1\over 2} \int{\rd^d p \over (2 \pi)^d}
   \log \left(p^2+ \chi + {\lambda \phi^2 \over 3} \right), 
 \ee
 up to an irrelevant constant.

 \begin{figure}[!ht]
\leavevmode
\begin{center}
\includegraphics[height=2cm]{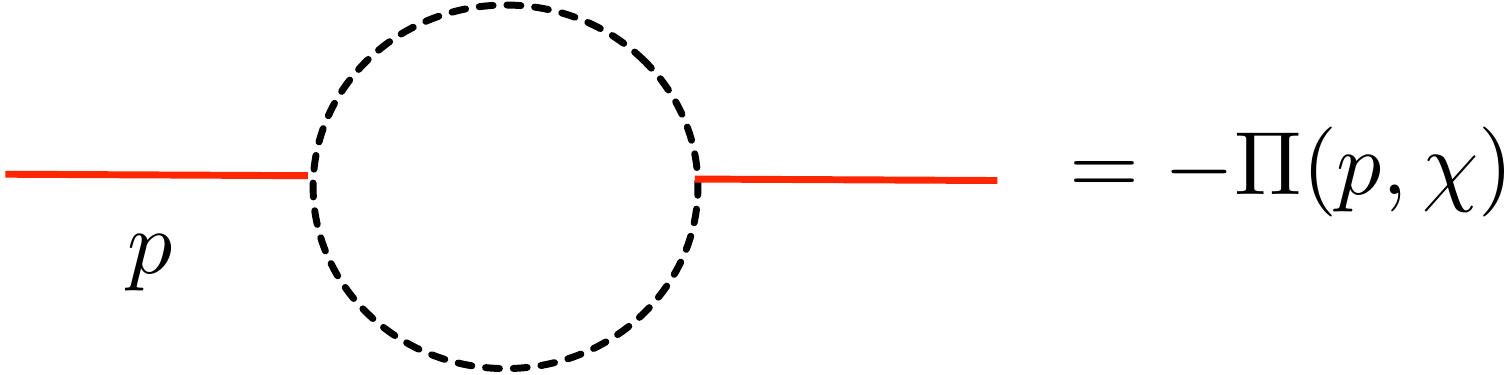}
\end{center}
\caption{The polarization loop, which gives the function $\Pi(p, \chi)$ introduced in (\ref{Pif}).}
\label{f-pol}
\end{figure} 

\begin{figure}[!ht]
\leavevmode
\begin{center}
\includegraphics[height=3cm]{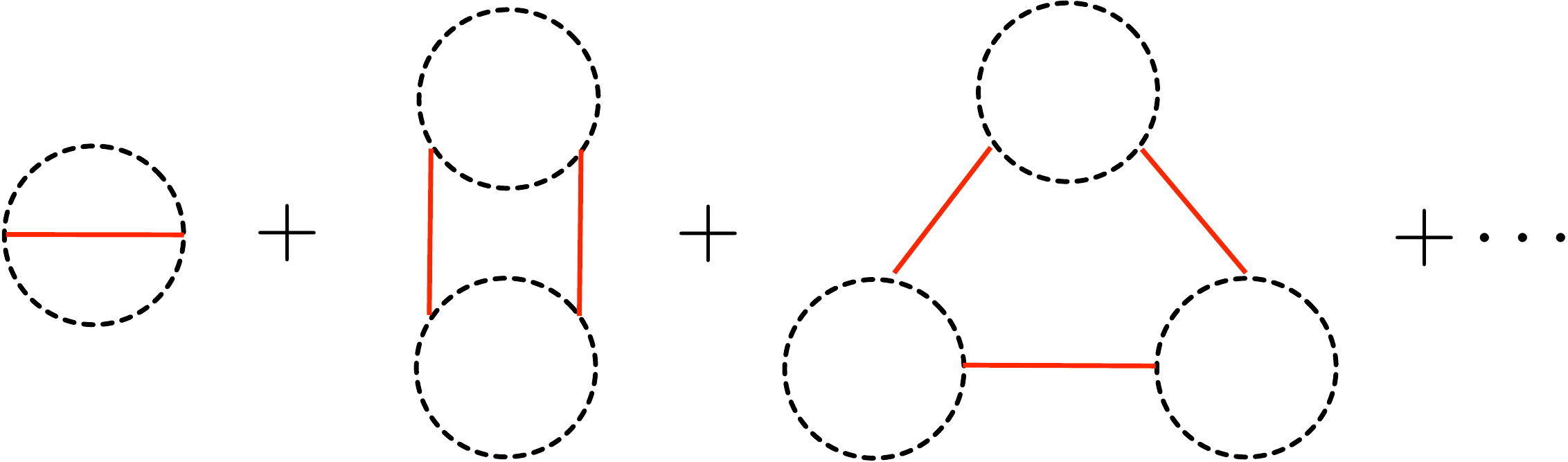}
\end{center}
\caption{Ring diagrams contributing to $V_{(1)} (\phi, \chi)$.}
\label{f-rings}
\end{figure}

 What other diagrams can contribute at this order? It turns out that the only contribution 
 comes from {\it ring diagrams}. The building block of ring 
 diagrams is the polarization loop $\Pi(p, \chi)$, which is shown in \figref{f-pol} and given by
\be
\label{Pif}
\Pi(p, \chi)= {1\over 2} \int  {\rd^2 q \over (2 \pi)^d} {1 \over (q^2 + \chi) ( (p+q)^2 +\chi)}. 
\ee
Note that this is of order one at large $N$. We can now join $n$ copies of this loop through $n$ 
propagators of the $\tilde \chi$ field, as shown in \figref{f-rings}. The sum of these ring diagrams gives the following 
contribution to the effective potential,  
 \be
 -\sum_{n \ge 1}{1\over 2n} \int {\rd^d p \over  (2 \pi)^d} (D_{\chi \chi} \Pi(p, \chi))^n =  -
 \sum_{n \ge 1}{1\over 2n} \int {\rd^d p \over  (2 \pi)^d} \left(-{\lambda\over 3} { (p^2+ \chi) \Pi(p,\chi)\over p^2 + \chi + {\lambda \phi^2  / 3}}\right)^n. 
 \ee
This can be summed in closed form to obtain 
\be
\label{subleading-corr-chi}
 {1\over 2} \int{\rd^d k \over (2 \pi)^d} \log\left[ {(k^2+ \chi) (1+\lambda \Pi(k,\chi)/3) + \lambda \phi^2/3 \over k^2 +\chi + \lambda \phi^2/3}\right]. 
\ee
Together with (\ref{half}) we obtain
\be
\label{subleading-corr}
V_{(1)}(\phi, \chi)= {1\over 2} \int{\rd^d k \over (2 \pi)^d} \log\left[ {(k^2+ \chi) (1+ \lambda \Pi(k, \chi)/3) + \lambda \phi^2/3 \over k^2 +\chi}\right]. 
\ee

We can now use these results to compute the ground state energy. By evaluating the effective potential at its minimum we find 
 \be
 \ba
 \label{potn}
 E&= N V_{(0)} (  \phi_{(0)}^2, \chi_{(0)})
  + {\partial V_{(0)} \over \partial \phi^2} (  \phi_{(0)}^2, \chi_{(0)})\phi^2_{(1)}
  + {\partial V_{(0)} \over \partial \chi} (  \phi_{(0)}^2, \chi_{(0)}) \chi_{(1)} \\
  &+ V_{(1)} (  \phi_{(0)}^2, \chi_{(0)})+ \CO\left({1\over N}\right). 
  \ea
 \ee
Due to the minimization conditions, the terms involving $\phi^2_{(1)}$, $\chi_{(1)}$ in the first line of (\ref{potn}) 
in principle vanish. One has to be careful, however, due to the regularization of IR divergences: since 
\be
{\partial V_{(0)} \over \partial \phi^2}=\chi_{(0)}= \epsilon^2,
\ee
we have to calculate the limit $\epsilon \rightarrow 0$ of the second term in the first line of (\ref{potn}) explicitly. 
By using the explicit expression for $\phi^2_{(1)}$ it can be checked that indeed
\be
 E= N V_{(0)} ( \phi_{(0)}^2, \chi_{(0)})+V_{(1)} ( \phi_{(0)}^2, \chi_{(0)})+ \CO\left({1\over N}\right). 
 \ee
 The leading term is indeed given by (\ref{leadinge0}). By using (\ref{subleading-corr}) and (\ref{leading-phi}), and taking into account our renormalization scheme, we obtain
\be
\CE_{(1)}= {1\over 8 \pi}+
{1\over 2m^2 } \int{\rd^2 k \over (2 \pi)^2} \log\left[ 1+
 { \lambda \over 3} {1\over k^2+m^2} \left( (k^2+ \epsilon^2) \Pi(k, \epsilon^2)+{1\over 4 \pi} \log \left( {\epsilon^2 \over m^2} \right)  \right) \right]. 
\label{elog}
 \ee
As emphasized in \cite{amit-kotliar}, the IR divergences for the ground state energy have to cancel order by order in the $1/N$ expansion. 
 It is now possible to check explicitly the absence of IR singularities in $\CE_{(1)}$. The polarization loop 
 can be computed explicitly (see e.g. Appendix A in \cite{mmbook}):
\be
\label{twopointf}
\Pi(p, \chi)  ={1\over 4 \pi {\sqrt{ p^2(p^2+ 4 \chi)}}} \log { {\sqrt{p^2 + 4\chi}} +{\sqrt{p^2}} \over   {\sqrt{p^2 + 4\chi}} -{\sqrt{p^2}}}.
\ee
In the limit $\chi=\epsilon^2 \rightarrow 0$ we find 
\be
\Pi(p, \epsilon^2) = {1\over 4 \pi p^2} \log\left({p^2 \over \epsilon^2} \right) + \CO(\epsilon^2), 
\ee
This is indeed IR divergent, but the divergence {\it cancels} against the 
contribution due to $\phi^2_{(0)}$. In the $\epsilon^2 \rightarrow 0$ limit, 
one finds 
\be
\label{e1-final}
\CE_{(1)}= {1 \over 8 \pi}+
{1\over 2m^2} \int{\rd^2 k \over (2 \pi)^2} \log\left[ 1+ { \lambda \over 12 \pi } {1\over k^2+m^2}  \log \left( {k^2 \over m^2} \right) \right].
 \ee
 This expression is IR finite but UV divergent, due to the linear term in $\lambda$. The UV divergence can be 
 removed by using our renormalization scheme. This is easier to do in the original representation (\ref{elog}), or by requiring that we 
  reproduce the two-loop result (\ref{e2loops}) at this order in the $1/N$ expansion. In this way, we obtain the manifestly
  finite, surprisingly simple answer for the subleading correction to the ground state energy in the $1/N$ expansion, 
  \be
  \label{tot-final}
\CE_{(1)}= {1 \over 8\pi} - {\pi \over 48} \gamma-\CI(\gamma), 
\ee
where $\CI(\gamma)$ is the integral 
\be
\label{int-gam}
\CI(\gamma)= {1\over 8 \pi } \int_0^\infty \rd x \left\{ \log\left[ 1+\gamma  { \log \left(x\right) \over x+1} \right] 
-\gamma { \log \left(x\right) \over x+1}  \right\}
\ee
and $\gamma$ is the dimensionless coupling
\be
\gamma = {\widehat \lambda \over 12 \pi}.
\ee
 It is useful to expand the function $\CI(\gamma)$ in powers of $\gamma$:
\be
\label{e1-series}
\CE_{(1)}\sim  {1\over 8 \pi}- {\pi \over 48} \gamma -  \sum_{n=0}^\infty c_n \gamma^{n+2}.  
 \ee
The coefficients $c_n$ are given by 
 \be
 \label{anco}
 c_n ={(-1)^n  \over  8 \pi (n+2)} \int_0^\infty \left( {\log(x) \over 1+x} \right)^{n+2} \rd x, \qquad  n \ge 0.  
 \ee
The integrals can be computed as derivatives of the beta function, as in a similar calculation in \cite{broad}: 
\be
\int_0^\infty { \rd x \over (x+ 1)^n} \left( \log (x) \right)^n=   (-1)^n {\rd^n  \over \rd z^n} B(1-z, n-1+z)\biggl|_{z=0}. 
\ee
Of course, the coefficients $e_{1, n+2}$ appearing in (\ref{erk1}), (\ref{erk2}) are given, up to a overall factor $(12 \pi)^{-n-2}$, by the $c_n$ in (\ref{anco}). 
We have then determined the leading contribution to the perturbative expansion (\ref{erk1}) when $N$ is large. 

What is the large order behavior of the $c_n$? Let us first note that their integrand involves the $n$-th power of the logarithm of the momentum, 
which is typical of renormalon diagrams \cite{beneke}. As $n$ grows, this integrand has larger and larger values near $k=0$, i.e. in the IR region, while the 
UV behavior as $k \rightarrow \infty$ is tamed by the denominator $(k^2+m^2)^{n}$. It is indeed easy to show that\footnote{In the next section we will derive a closed formula for the $c_n$, 
as well as a precise asymptotic expansion for them.} 
\be
c_n \sim {n! \over 8\pi } , \qquad n \gg 1.
\ee
Therefore, the first non-trivial contribution to the ground state energy in the $1/N$ expansion is 
given by a factorially divergent series in the coupling constant. The factorial 
growth is due to the integration over momenta in the IR region, and leads to a singularity in the Borel plane 
of the dimensionless coupling constant $\widehat g$. The singularity is located, at large $N$, at 
\be
\label{b-sing}
\zeta= {12 \pi \over N }. 
\ee
Therefore, this is an IR renormalon. It leads to a non-perturbative ambiguity in this theory, characterized by the exponentially small scale
\be
\label{ambig}
\exp \left(- {12 \pi \over N \widehat g}\right). 
\ee
Since the singularity is in the positive real axis, the perturbative expansion in this theory is not Borel summable, for $N>1$. This is in contrast to the case $N=1$, where 
the series is Borel summable \cite{eckmann,blgzj-ins, serone1, serone2}. It is interesting to note that the argument in \cite{blgzj-ins, serone1, serone2} 
for Borel summability can not be applied in the case of two-dimensional theories with a continuum of degenerate vacua, which is precisely the case we are looking at. 

Let us make some additional comments on this result: 

\begin{enumerate}

\item
Although the coupling constant $g$ does not get renormalized, the mass does, and this leads to a 
renormalization of the dimensionless coupling constant $\widehat g$. The corresponding beta function at one-loop is 
\be
\beta(\widehat g)= -{N+2 \over 6 \pi} \widehat g^2. 
\ee
The coefficient appearing in the exponent of (\ref{ambig}) can then be interpreted as twice 
the coefficient of this beta function, at large $N$. This is similar to what happens with standard IR renormalons
in an asymptotically free theory. 

\item A similar calculation can be done in the theory (\ref{lag}) but in three dimensions, which is also 
super-renormalizable. In this case, the symmetry remains broken quantum-mechanically \cite{cjp} and the classical vacuum is the ``true" vacuum at 
weak coupling. The $1/N$ 
correction to the ground state energy leads to a series which grows 
only exponentially and has a finite radius of convergence, which is the expected behavior in the absence of 
renormalons (see e.g. \cite{mmbook,mmlargen}). This supports our interpretation of the IR renormalon in (\ref{e1-series}) as a manifestation of the Coleman--Mermin--Wagner theorem.

\item As in the case of QED or QCD renormalons, the IR renormalon at (\ref{b-sing}) is associated to bubble-like diagrams, 
which in the Hubbard--Stratonovich form of the theory (\ref{lag-chi}) are the ring diagrams of \figref{f-rings}. In order to cancel IR divergences, 
however, one also has to add the one-loop contribution (\ref{half}). In the original formulation of the theory, the factorially divergent behavior is not due to a single class of diagrams but is the result of different types of diagrams that have to be combined in order to obtain an IR finite result. The large $N$ expansion 
finds the right combination of diagrams for us.  
 
\item The original integral form (\ref{tot-final}) provides a resummation of the power series (\ref{e1-series}), but it gives a {\it complex} result, 
for any value of $\lambda$. The imaginary part is of order (\ref{ambig}).  
This is yet another manifestation of the lack of Borel summability of the original series, as we will make more precise in the next section. 

\item It is interesting to note that the non-perturbative scale (\ref{ambig}) shows up when this model is studied on a two-dimensional 
Anti-de Sitter space \cite{io, pietro}. In AdS$_2$ the Coleman--Mermin--Wagner 
theorem can be evaded and, if the AdS$_2$ radius $R$ is sufficiently small, there is a phase where the $O(N)$ symmetry is 
spontaneously broken. The boundary separating this phase from the phase with 
unbroken symmetry is given by \cite{io}
\be
R M = \exp \left(- {12 \pi \over N \widehat g}- \gamma_E\right), 
\ee
where $M$ is a renormalization mass scale (we note that in \cite{io} the UV divergences are regularized by a cut-off). This involves 
precisely the non-perturbative ambiguity (\ref{ambig}).

\item In \cite{serone2}, complex instantons were found for the theory (\ref{z2-version}) 
with $\mu^2<0$ and with a single scalar field (i.e. $N=1$). These solutions to the classical EOM can be trivially embedded in the 
$O(N)$ theory we have studied, and we expect them to lead to singularities in the Borel plane. 
Singularities associated to complex instantons do not obstruct Borel summability, but they might 
control the leading large order behavior of the perturbative series for some values of $N$, 
similarly to what happens in \cite{voros-quartic}. 

\item There might be additional instanton singularities in this theory, and one could ask 
whether the singularity (\ref{b-sing}) 
can be associated to an instanton. Usually, factorial growth in the loop expansion at a fixed order in the 
$1/N$ expansion can not be justified by instanton singularities (see e.g. \cite{fkw,mmlargen}). In addition, it is easy to see that any real solution 
to the Euclidean EOM has a negative real action, 
\be
\label{s-rel}
S=-{g \over 4!} \int \rd^d x \, \bPhi^4(x), 
\ee
therefore an instanton configuration leading to the singularity at (\ref{b-sing}) has to 
be complex but lead to a positive, real action. Moreover, this action should 
scale as $1/N$ for large $N$. In some models, and after a twisted compactification, 
such scaling can be obtained through fractional
 instantons \cite{a-unsal-long, dunne-unsal-cpn,cherman-dorigoni-dunne-unsal, cdu, 
 misumi-1,du-on, misumi-2}. However, in this case, and in infinite volume, an instanton interpretation of 
the singularity (\ref{b-sing}) seems difficult to achieve. 

\item The structure of (\ref{tot-final}) is very similar to the $1/N$ correction to the ground state energy of the attractive Gaudin--Yang model with $N$ components, calculated 
in \cite{mr-long}. In both cases, ring diagrams diverge factorially and they are resummed by a logarithm which gives an exponentially small 
imaginary part. In the case of the Gaudin--Yang model, this imaginary part is a manifestation of the Cooper instability.  

\end{enumerate}

\subsection{Resurgent analysis}

As we have seen in the previous section, the series (\ref{e1-series}), which is obtained from ring diagrams, 
diverges factorially and leads to a singularity in the Borel plane. Factorially divergent series arising in the large 
$N_f$ limit of QED (or the large $\beta_0$ limit of QCD) can sometimes 
be analyzed in great detail, and their Borel transforms computed in closed form, see \cite{beneke} for examples. 
In this section we will perform such an analysis for (\ref{e1-series}) by using the theory of resurgence 
(for this example, we will only need the tools presented in \cite{mmbook}). 

Let us consider the formal power series appearing in (\ref{e1-series}), 
\be
\varphi(\gamma)= \sum_{n=0}^\infty c_n \gamma^n, 
\ee
as well as its Borel transform 
\be
\label{bt}
\widehat \varphi(\zeta)= \sum_{n=0}^\infty {c_n \over n!} \zeta^n. 
\ee
This Borel transform has singularities on the positive real axis, so the conventional Borel resummation is not well-defined. However, we can define the lateral Borel resummations as
\be
\label{lateralborel-theta}
s_{\pm} (\varphi)(z)=z^{-1} \int_{\CC_{\pm}} \rd \zeta \, \re^{-\zeta/z} \widehat \varphi (\zeta), 
\ee
where $\CC_\pm$ are integration paths slightly above (respectively, below) the positive real axis. These lateral Borel resummations have an imaginary piece 
which, according to the theory of resurgence, can be obtained by an appropriate resummation of a formal trans-series, involving both the coupling $\gamma$ and the 
exponentially small coupling $\re^{-1/\gamma}$. We expect this 
trans-series to be of the form 
\be
\label{ts}
\sum_{\ell=1}^\infty C_\ell \re^{-\ell/\gamma} \gamma^{-b_\ell} \varphi_{\ell}(\gamma), 
\ee
where $C_\ell$ are constants (sometimes called the trans-series parameters) and 
\be
\varphi_\ell(z)= \sum_{n \ge 0} a_{\ell,n} \gamma^n 
\ee
is a, in general divergent, formal power series in $\gamma$. 

It turns out that in this example the trans-series (\ref{ts}) can be computed in a simple way. The reason is the following. 
The resummed energy can be written in terms of the integral $\CI(\gamma)$ in (\ref{int-gam}). As we mentioned in the previous section, 
this integral has an imaginary part which is easy to calculate. For any value of $\gamma$, the argument of the logarithm becomes negative for 
\be 
0<x<x(\gamma), 
\ee
where $x(\gamma)$ is the solution to the equation 
\be
\label{zero-int}
1+\gamma  { \log \left(x\right) \over x+1} =0. 
\ee
Along the interval $[0, x(\gamma)]$, the integrand of (\ref{int-gam}) has a constant imaginary part, given by $\pm \ri/8$, where the 
sign depends on a choice of sign for $\log(-1)=\pm \pi \ri$. 
The imaginary part of $\CI(\gamma)$ is then given by
\be
 {\rm Im}\, \CI(\gamma)= \pm {\ri \over 8} x(\gamma). 
\ee
We expect this imaginary part to agree (up to an overall factor $\gamma^2$) with the imaginary part of the lateral Borel resummations (\ref{lateralborel-theta}). 
On the other hand, an explicit solution to (\ref{zero-int}) can be found by using Lambert's function $W(x)$:
\be
\label{x-ts}
x(\gamma)=\gamma W\left( \gamma^{-1} \re^{-1/\gamma} \right)= \sum_{\ell=1}^\infty {\ell^{\ell-1} \over \ell!} (-1)^{\ell-1} \gamma^{1-\ell} \re^{-\ell/\gamma}. 
\ee
From this we can read the trans-series (\ref{ts}) 
associated to $\varphi(\gamma)$: it has 
\be
b_\ell= \ell+1, \qquad \varphi_\ell(\gamma)={1\over 8}  {\ell^{\ell-1} \over \ell!} (-1)^{\ell-1}. 
\ee
In this case, the series $\varphi_\ell(\gamma)$ truncate to a single coefficient (this truncation seems to be typical of trans-series 
appearing in large $N$ renormalon calculations). 
Note in addition that, as a series in $\re^{-1/\gamma}/\gamma$, the trans-series has a finite radius of convergence. This also happens in 
other examples (see e.g. \cite{costin,multi-multi}). One can verify numerically that the integral $\CI(\gamma)$ agrees with the lateral Borel resummation, i.e. 
\be
\CI(\gamma) =\gamma^2 s_\pm (\varphi)(\gamma), 
\ee
where the choice of lateral resummation corresponds to the choice of sign in the imaginary part of $\CI(\gamma)$. 

 \begin{figure}[!ht]
\leavevmode
\begin{center}
\includegraphics[height=3cm]{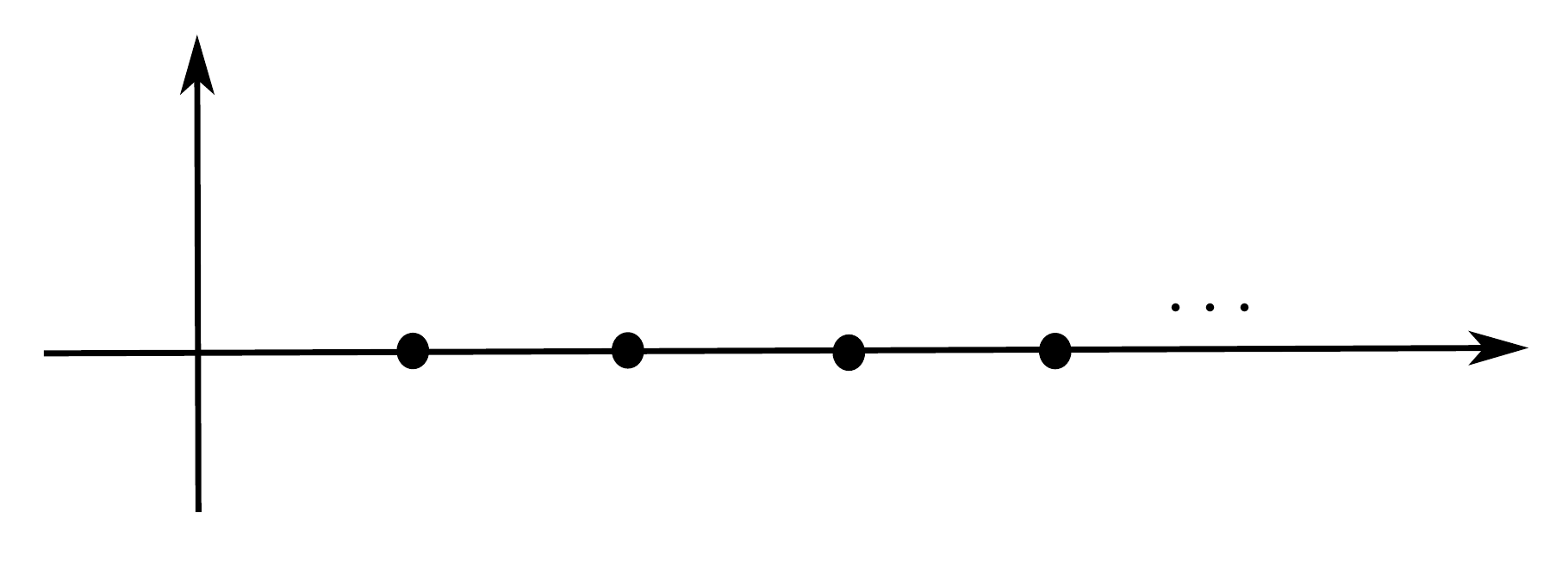}
\end{center}
\caption{The Borel transform (\ref{bt}) has poles of order $\ell+1$ at all positive integers $\ell=1, 2, \cdots$. The $\ell$-th singularity leads to an exponentially small 
correction $\exp\left(-12 \pi \ell/(N \widehat g)\right)$ in the trans-series.}
\label{2ds-fig}
\end{figure}

It follows from the general theory of resurgence that the trans-series contains information about the singularities 
of the Borel transform of the original series. In particular, a pole of order $k$ at $\zeta=A$ in the 
Borel transform $\widehat \varphi(\zeta)$, with coefficient $a$, leads to a term in the trans-series of the form 
\be
{ \pi a (-1)^{k-1} \over (k-1)!} {\re^{-A/\gamma} \over \gamma^k}.
\ee
In our case, we find poles at all positive integers $\zeta=\ell$, of order $\ell+1$. More precisely, we can write
\be
\label{poles-reg}
\widehat \varphi(\zeta)= {1\over 8 \pi} \sum_{\ell=1}^\infty {\ell^{\ell-1} \over (\zeta-\ell)^{\ell+1}} + \text{regular}. 
\ee
This gives the precise positions of all the IR renormalon singularities, which we show schematically in \figref{2ds-fig}. 
We note that there are no other singularities in the Borel plane. In particular, 
there are no UV renormalons. The regular part in (\ref{poles-reg}) can be guessed by using the fact that the coefficients $c_n$ only involve zeta 
functions evaluated at {\it even} integers. We then conjecture the following 
exact expression, 
\be
\label{bt-ex}
\widehat \varphi(\zeta)= {1\over 8 \pi} \sum_{\ell=1}^\infty \left\{ {\ell^{\ell-1} \over (\zeta-\ell)^{\ell+1}} + (-1)^{\ell-1} \ell^{-\ell-1}  (\zeta+\ell)^{\ell-1} \right\}, 
\ee
which we have checked by expanding around $\zeta=0$ at very large order. Interestingly, we can now reconstruct the original perturbative coefficients in (\ref{bt}) by the Cauchy formula
\be
{c_k \over k!} = {1 \over 2\pi \ri} \oint_{\CC_0}   {\widehat \varphi (\zeta)  \over \zeta^{k+1}} \rd \zeta, 
\ee
where $\CC_0$ is a small circle around $\zeta=0$. We deform the contour $\CC_0$ to pick all the poles in the positive real axis, 
plus a contour at infinity $\CC_\infty$. We obtain in this way 
\be
\label{c-deff}
c_k = -k! \sum_{\ell=1}^\infty  {\rm Res}_{\zeta=\ell } \left( {\widehat \varphi (\zeta)  \over \zeta^{k+1}}\right) + {k! \over 2\pi \ri} \oint_{\CC_\infty}   {\widehat \varphi (\zeta)  \over \zeta^{k+1}} \rd \zeta.
\ee
The sum in the r.h.s. of (\ref{c-deff}) gives the contribution of Borel singularities to the large order behavior of the perturbative series. 
By using (\ref{bt-ex}) we find 
\be
\label{lo-ser}
{1\over 8 \pi} \sum_{\ell=1}^\infty \ell^{-k-\ell-1} \Gamma(k+ \ell+1)  {(-1)^{\ell-1} \ell^{\ell -1}\over \ell!}={1\over 8 \pi} \sum_{\ell=1}^\infty (-1)^{\ell-1} \ell^{-k-2} (1+\ell)_k, 
\ee
where $(\alpha)_\beta$ is the Pochhammer symbol. The expression in the l.h.s. has the standard form of a resurgent asymptotic formula. The contribution of the 
integral around infinity in (\ref{c-deff}) gives
\be
{1\over 8 \pi} \sum_{\ell=1}^\infty (-1)^{\ell-1} {(k-\ell)(k-1-\ell) \cdots (1-\ell) \over (-\ell)^{k+2}}, 
\ee
and we obtain
\be\label{cksuma}
c_k= {1\over 8 \pi} \sum_{\ell=1}^\infty (-1)^{\ell-1} \ell^{-k-2} \left( (1+\ell)_k + (-1)^k (1-\ell)_k \right). 
\ee
This can be written even more explicitly by expressing the
Pochhammer symbols in terms of Stirling numbers of the first kind $s(k,n)$. We find, 
\be
c_k= {1\over 4 \pi}\sum_{t=0}^{[{k+1 \over2}]} s(k+1, k+1-2t) \left(1- 2^{-1-2t}\right) \zeta(2t+2). 
\ee
This provides a much more explicit expression for the coefficients of the perturbative series than the original integral formula (\ref{anco}). Therefore, in this case, 
knowledge of the trans-series eventually leads to a better understanding of the perturbative series. Interestingly, the appearance of zeta functions in the perturbative 
coefficients is closely related to the sum over the singularities of the Borel transform at positive integer points. 

Let us note that, if we are interested in an asympotic expansion of $c_k$ at large $k$, the second term in the r.h.s. of (\ref{cksuma}) (which comes from the 
contour integral at infinity) does {\it not} contribute, since $(1-\ell)_k=0$ for any $k\ge l$. Therefore, as an asymptotic expansion, we have
\be
c_k \sim {1\over 8 \pi} \sum_{\ell=1}^\infty \ell^{-k-\ell-1} \Gamma(k+ \ell+1)  {(-1)^{\ell-1} \ell^{\ell -1}\over \ell!}, \qquad k \gg 1. 
\ee
In particular, the leading asymptotics comes from the first singularity at $\ell=1$,  
\be
c_k \sim {1\over 8 \pi} \Gamma(k+2), \qquad k \gg 1, 
\ee
with no corrections of order $1/k$ (the corrections to this formula are exponentially small, of order $2^{-k}$).

\subsection{Generalization to an arbitrary potential} 

The result (\ref{tot-final}) for the ground state energy at subleading order in $1/N$ can be easily generalized to a scalar theory with an arbitrary potential $U(x)$, described by the Lagrangian
\be
\label{gen-lag}
\CL(\bPhi)= {1\over 2} \partial_\mu \bs{\Phi}\cdot \partial^\mu \bs{\Phi}-N U\left( \bPhi^2/N \right). 
\ee
At the classical level, we have spontaneous symmetry breaking if the equation 
\be
U'(\phi^2)=0 
\ee
has non-trivial positive solutions. We will assume that this is the case, and we will denote by 
\be
\phi^2_\star
\ee
the value of $\bPhi^2/N$ at the minimum. At tree level we find a massive particle $\xi$ with square mass 
\be
m^2= 4 \phi_\star^2 U''(\phi_\star^2)
\ee
and $N-1$ Goldstone bosons $\bs{\eta}$. 

The calculation of the effective potential of (\ref{gen-lag}) at large $N$ was done in \cite{schnitzer}, 
while the next-to-leading correction 
was calculated in \cite{rembiesa}. They generalize the quartic case considered in \cite{cjp,root} as well as the sextic case considered 
in \cite{townsend1,townsend2}. We will present a simpler derivation, by combining ingredients from \cite{root, zjbook, onsg}. 
The first step, following \cite{zjbook}, is to introduce two new scalar fields in the path integral, $X$ and $\Sigma$, 
by the following delta function trick:
\be
1= \int {\cal D} \Sigma \, \delta( \Sigma- \bPhi^2/N)= \int {\cal D} \Sigma \, {\cal D} X\, \exp \left\{ \ri N X/2 \left( \Sigma- \bPhi^2/N \right) \right\}. 
\ee
This leads to an equivalent theory with Lagrangian
\be
\CL(\bPhi, X, \Sigma)= {1\over 2} \partial_\mu \bs{\Phi}\cdot \partial^\mu \bs{\Phi}-{X \over 2} \bPhi^2 - N U(\Sigma) + {N \Sigma X \over2}. 
\ee
As usual we expand around the constant configuration
\be
\ba
\Phi_1(x) &= {\sqrt{N}} \phi+\xi (x), \qquad \Phi_j(x)= \eta_{j-1}(x), \qquad j=2, \cdots, N,\\
 X(x)&= \chi + \tilde \chi (x)/{\sqrt{N}}, \qquad \Sigma= \sigma+ \tilde \sigma (x)/{\sqrt{N}}. 
 \ea
\ee
We then obtain 
\be
\ba
\CL(\bPhi, X, \Sigma)&=-N \left( U(\sigma)-{\sigma \chi \over 2}+ {\chi \phi^2 \over 2} \right)+  {1\over 2}  \partial_\mu \bs{\eta} \cdot \partial^\mu \bs{\eta} -{1\over 2} \chi  \bs{\eta}^2 \\ 
&+{1\over 2} \partial_\mu \xi \cdot \partial^\mu \xi -{1\over 2} \chi \xi^2 - \tilde \chi \xi \phi  -{1\over 2} U''(\sigma) \tilde \sigma^2 +{1\over 2} \tilde \chi \tilde \sigma \\
&-{1\over 2{\sqrt{N}}} \tilde \chi \xi^2 -{1\over 2{\sqrt{N}}}  \tilde \chi \bs{\eta}^2.
\ea
\ee
As in previous calculations, it is more convenient to rotate to Euclidean signature. The $N-1$ scalars lead to a one-loop correction 
\be
{N-1 \over 2} \int {\rd^d k \over (2 \pi)^d} \log (k^2+ \chi), 
\ee
and the final result for the effective potential at leading order in the $1/N$ expansion is 
\be
V_{(0)}(\sigma, \chi, \phi)=  U(\sigma)-{\sigma \chi \over 2}+ {\chi \phi^2 \over 2}+ {1 \over 2} \int {\rd^d k \over (2 \pi)^d} \log (k^2+ \chi). 
\ee
We can compare this result to the one obtained in \cite{schnitzer,rembiesa}. To do this, we solve for $\chi$, $\sigma$, by using the minimization conditions
\be
\ba
{\partial V_{(0)}\over \partial \sigma}&= U'(\sigma)-{\chi \over 2}=0, \\
{\partial V_{(0)}\over \partial \chi}&= {\phi^2 \over2}-{\sigma \over 2}+{1 \over 2} B_1(\chi), 
\ea
\ee
where
\be
B_1(\chi)= \int {\rd^d k \over (2 \pi)^d} {1 \over k^2+ \chi}. 
\ee
We then solve 
\be
\label{gp-min}
\sigma= \phi^2+ B_1(\chi), \qquad \chi= 2 U'\left(\phi^2 + B_1(\chi) \right). 
\ee
The second equation defines $\chi$ implicitly. We conclude that 
\be
V_{(0)}(\phi^2)= U\left( \phi^2 + B_1(\chi)  \right)-B_1(\chi) U'\left( \phi^2 + B_1(\chi)  \right)+  {1 \over 2} \int {\rd^d k \over (2 \pi)^d} \log (k^2+ \chi). 
\ee
This is precisely the result obtained in \cite{schnitzer,rembiesa}. 

We can now calculate the next-to-leading term in $1/N$ as we did in the quartic case, following \cite{root}. The inverse propagator for the three fields, $\xi$, $\tilde \chi$, $\sigma$, is given by the matrix
\be
D^{-1}= \begin{pmatrix} p^2+ \chi & \phi & 0 \\ \phi & 0 & -{1\over 2} \\
0 & -{1\over 2} & U''(\sigma) \end{pmatrix}. 
\ee
From this we obtain the $\tilde \chi \tilde \chi$ propagator, 
\be
D_{\tilde \chi \tilde \chi}(p)= -4 U''(\sigma) {p^2 + \chi \over p^2+ \chi+ 4 U''(\sigma)\phi^2}. 
\ee
The final result is
\be
\label{subleading-gen}
 V_{(1)}(\sigma, \chi, \phi)= {1\over 2} \int{\rd^d k \over (2 \pi)^d} \log\left[ {(k^2+ \chi) (1+ 4 U''(\sigma) \Pi (k,\chi)) +4 U''(\sigma) \phi^2  \over k^2 +\chi}\right], 
\ee
in agreement with \cite{townsend2,rembiesa}. 

Let us now calculate the vacuum energy at this order. The solution of (\ref{gp-min}) is 
\be
\sigma=\phi^2+ B_1(\chi) = \phi_\star^2+ \CS(\chi), 
\ee
where $\CS(\chi) \rightarrow 0 $ when $\chi \rightarrow 0$. The function appearing in (\ref{subleading-gen}) then reads, as $\chi \rightarrow 0$, 
\be
\ba
& (k^2+ \chi) (1+ 4 U''(\sigma) \Pi (k,\chi)) +4 U''(\sigma) \phi^2\\
&= k^2+ m^2 + 4 U''(\phi^2_\star) \left\{ (k^2+ \chi) \Pi(k, \chi)+ {1\over 4 \pi} \log\left( {\chi \over m^2}\right) \right\}+ \CO(\chi), 
\ea
\ee
where we have renormalized $B_1(\chi)$ by using the prescription (\ref{reg-int}) (this is equivalent, at large $N$, to renormalizing the coefficients of the 
potential, which is in turn equivalent to normal-ordering \cite{schnitzer}). Like before, IR divergences cancel, and we find that $\CE_{(1)}$ involves the series
\be
- \sum_{n \ge 0} c_n \left( {1\over 4 \pi \phi_\star^2} \right)^{n+2}, 
\ee
where the coefficients $c_n$ are the same ones (\ref{anco}) appearing in the theory with a quartic potential. The non-perturbative ambiguity is now of the form 
\be
\exp\left(-4 \pi \phi_\star^2\right). 
\ee

\sectiono{On the Lieb--Liniger model}
\label{sec-ll}

A Bose gas at zero temperature with chemical potential $\mu$ and a repulsive $\delta$-function interaction is described by the following action, 
\begin{equation}
S(\Phi)=\; 
\int \rd t\,\int \rd^D x\;
\left\{ \Phi^\dagger 
       \left( \ri\partial_t + {\nabla^2 \over 2m} + \mu \right) \Phi
        - {1 \over 4} g\, (\Phi^\dagger \Phi)^2 \right\},  
\label{actpsi}
\end{equation}
where $\Phi$ is a complex scalar field. The grand potential $\Omega(\mu)$ of this gas can be calculated by computing first the effective potential $\CV(\mu, \phi)$, 
as a function of the vev of $\Phi$, 
$\phi$. The value of $\phi$ is then fixed by the minimization condition 
\be
 \label{Omega-min}
     {\partial \CV \over\partial \phi}(\mu,\phi) =0. 
   \end{equation}
Finally, $\Omega(\mu)$ is obtained by
evaluating $\CV(\mu,\phi)$ at this minimum. 

When $D=1$, the model described by (\ref{actpsi}) is integrable and known as the Lieb--Liniger model \cite{ll}. In this case $\Omega(\mu)$ can be calculated with the Bethe ansatz. It has a 
perturbative expansion in powers of the 
coupling constant, with the following structure
 \be
 \label{pseries}
 { g \over \mu^2} \Omega(\mu) = -\sum_{n \ge 0} c_n \xi^{n}, 
 \ee
where
\be
\xi= {g \over 4 \mu^{1/2}}. 
\ee
Up to order $g$, one finds \cite{popov, tw-ll, mr-ll}
\be
\label{bethe}
\Omega(\mu)=-{\mu^2 \over g}-{\sqrt{2} \over 3 \pi} \mu^{3/2} +{g \mu \over 4}\left( {1\over 12}- {1\over \pi^2} \right)+ \cdots
\ee
In \cite{mr-ll} we calculated the first fifty coefficients of the expansion (\ref{pseries}), and we found numerically the following asymptotics 
\be
\label{om-as}
c_n \sim A^{-n} n!, \qquad A= 4 {\sqrt{2}} \pi.
\ee
In particular, the perturbative series is not Borel summable. 

As we mentioned in the Introduction, there are two possible sources for the behavior (\ref{om-as}): it could be due to an instanton configuration, 
and then $A>0$ is identified with the action of an instanton, or it could be due to renormalon diagrams. Let us first consider instantons. 
These are solutions of the Euclidean equations of motion (EOM) for the action (\ref{actpsi}), and they are given by (see e.g. \cite{no})
\be
\ba
\left\{ {\partial \over \partial \tau} -{\nabla^2 \over 2m} - \mu + {g \over 2} |\Phi|^2  \right\} \Phi&=0, \\
\left\{ -{\partial \over \partial \tau} -{\nabla^2 \over 2m} - \mu + {g \over 2} |\Phi|^2  \right\} \Phi^\dagger&=0. 
\ea
\ee
 It is easy to see that any solution to the above equations leads to an Euclidean action given by 
\be
-{g\over 4} \int \rd \tau \rd^D x  \, (\Phi^\dagger \Phi)^2, 
\ee
similarly to (\ref{s-rel}). 
Therefore, if $\Phi^\dagger$ and $\Phi$ are complex conjugates, the action is negative and such an instanton, even if it exists, 
can not explain the Borel singularity in the positive real axis. In principle, one could consider more general instanton solutions in which $\Phi$ is not the complex conjugate of $\Phi^\dagger$, but we have not 
found any solution to the EOM which leads to a positive, real action\footnote{We have benefitted from many discussions with Peter Wittwer on this issue.}. 
It is therefore natural to suspect that the Borel 
singularity associated to the large order behavior (\ref{om-as}) is rather a renormalon singularity. 

In fact, the Lieb--Liniger model is in many ways a non-relativistic avatar of the $O(N)$ model that 
we studied in the previous section. For example, in $D=1$, the field $\Phi$ in (\ref{actpsi}) cannot have a vev quantum-mechanically, 
due to the non-relativistic version of the Coleman--Mermin--Wagner theorem. However, in the standard perturbative approach to the interacting Bose gas, we expand around 
such a ``false vacuum". As in the example studied by Jevicki and reviewed in section \ref{sec-2loop}, we expect to have IR divergences which eventually cancel in the calculation of observables. In view of this, 
it is natural to conjecture that the perturbation theory of the Lieb--Liniger mode has an IR renormalon, as in the $O(N)$ model, explaining in this way the large order behavior (\ref{om-as}). 

We will now analyze the perturbative structure of the Lieb--Liniger model up to two-loops. Such an analysis 
was performed long ago by using a collective variable formulation \cite{taka-nnl} and Popov's hydrodynamic formalism \cite{popov} (see \cite{pastukhov}). 
As emphasized in \cite{jevicki2}, this formalism is manifestly IR finite. However, it has other important drawbacks; for example, it contains an infinite number of interaction vertices. Therefore, 
and specially in view of further explorations of the perturbative structure, we will use the field-theoretic approach of \cite{b-nieto}, which considered the three-dimensional case $D=3$, and 
we will study the model in $D=1$ (the case $D=2$ was analyzed in \cite{2andersen}, and a useful review can be found in \cite{andersen-rev}). Many aspects of this analysis are independent of the dimension, 
and we refer to these references for further details. 

\begin{figure}[!ht]
\leavevmode
\begin{center}
\includegraphics[height=3.5cm]{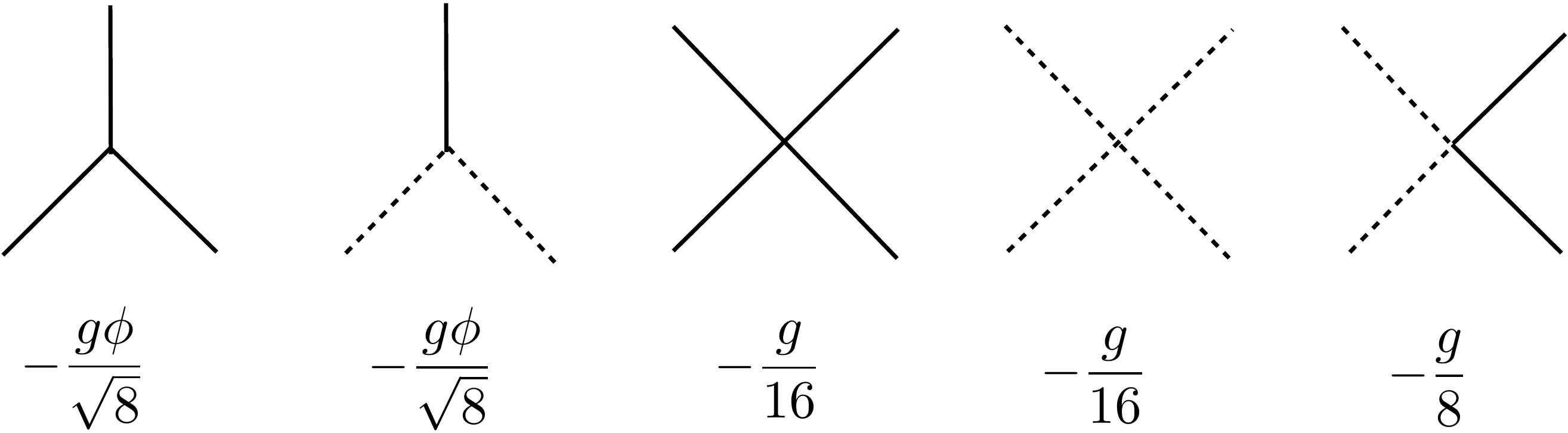}
\end{center}
\caption{The vertices for the interactions between $\eta$, $\xi$ fields.}
\label{vertex-bose}
\end{figure}

We first expand the field $\Phi$ around a constant configuration $\phi$ which breaks the global $U(1)$ symmetry:
\be
\label{psi-cart}
\Phi= \phi+ {\xi + \ri \eta \over {\sqrt{2}}}, 
\ee
where $\xi$, $\eta$ are real fields and $\phi$ is a real positive constant. The action becomes
\begin{equation}
S(\Phi) \;=\;  S(\phi) 
\;+\; S_{\rm free}(\xi,\eta) \;+\; S_{\rm int}(\phi,\xi,\eta)  \,.
\label{S-decomp}
\end{equation}
Here, 
\be
\label{sv}
S(\phi)= V T \left( \mu \phi^2 -{1\over 4} g \phi^4\right)
\ee
is the action evaluated at $\Phi=\phi$. The free part of the action consists of the terms quadratic in 
$\xi$ and $\eta$: 
\be
\ba
S_{\rm free}(\xi,\eta) &=  \int \rd t\,  \rd^D x \Bigg\{ 
{1 \over 2} \left( \eta \dot{\xi}- \xi \dot{\eta} \right)+ {1 \over 4 m} \xi \left( \nabla^2 
        - 2 mg \phi^2 + X \right) \xi 
       +{1 \over 4 m}\eta \left( \nabla^2 
        + X \right) \eta\Bigg\},
\label{S-free}
\ea
\ee
where 
\begin{equation}
      X = 2m\left( \mu - {1\over 2} g \phi^2 \right) \,.
   \label{X-def}
   \end{equation}
From $S_{\rm free}$ we can read the propagator for the fields 
$\xi$ and $\eta$: 
\be
D(\omega,k,\phi) =
{\ri \over  \omega^2 - \varepsilon^2(k,\phi) + \ri\epsilon}
\begin{pmatrix}
        (k^2 - X)/2m & - \ri \omega       \\
        \ri \omega    & (k^2 + 2mg \phi^2 - X)/2m 
        \end{pmatrix},
\label{propagator}
\end{equation}
where
\be
\label{epsi}
\varepsilon^2(k,\phi)= {1\over 4m^2} (k^2-X)\left(k^2 +2mg \phi^2-X\right). 
\ee
 \begin{figure}[!ht]
\leavevmode
\begin{center}
\includegraphics[height=2cm]{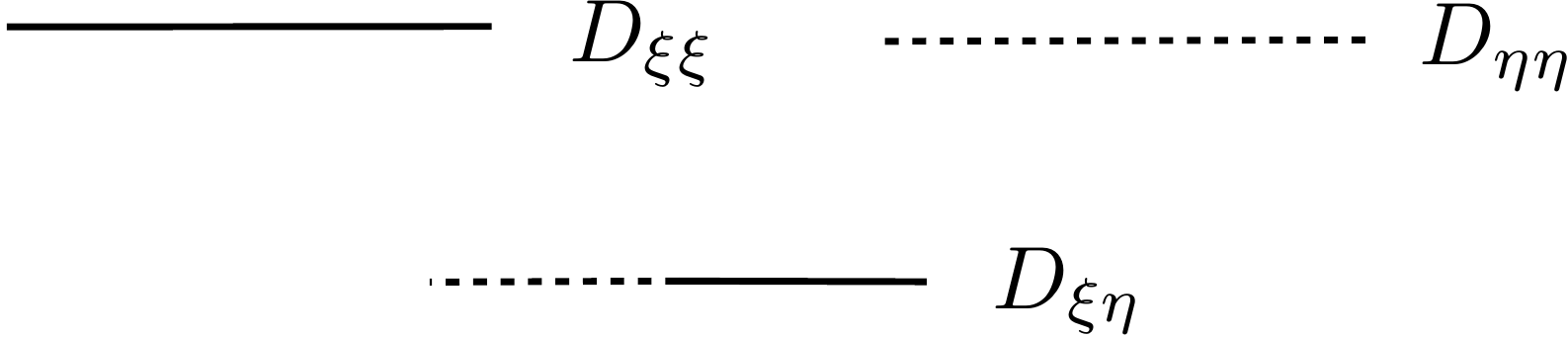}
\end{center}
\caption{The propagators for the $\eta$, $\xi$ fields.}
\label{feynman-bose}
\end{figure} 
The diagonal elements of the propagator matrix
(\ref{propagator}) are represented by solid lines for $\xi$
and dashed lines for $\eta$. The off-diagonal elements are represented by a
line that is half solid and half dashed, as illustrated in \figref{feynman-bose}. Finally, the interaction part of the action is given by
\begin{equation}
S_{\rm int}(\phi,\xi,\eta)=\; \int \rd t \int \rd^D x \Bigg\{ 
{\phi X\over \sqrt{2} m} \xi
\;-\;  {g\phi \over \sqrt{8}} \xi \left( \xi^2 + \eta^2 \right) 
\;-\;  {g \over 16}\left(  \xi^2 + \eta^2 \right)^2 
\Bigg\} \,.
\label{Sint-cart}
\end{equation}
It leads to the vertices shown in \figref{vertex-bose}.

Let us now calculate the loop expansion of $\CV(\mu, \phi)$. The tree-level contribution can be obtained from $S(\phi)$:
\begin{equation}\label{ome0}
  \CV_0(\mu,\phi) = - \mu \phi^2 + {1\over 4} g \phi^4 \, .
\end{equation}
The one-loop contribution is
\begin{equation}
      \CV_1(\mu,\phi) = {\ri\over 2}
      \int {\rd^D k\over (2\pi)^D}
      \int {\rd\omega\over 2\pi} \log \det D(\omega,k,\phi) =  {1\over 2}\int {\rd^D k\over (2\pi)^D}\,\varepsilon(k,\phi). 
\end{equation}
One-loop calculations in this theory involve the integrals 
\be
\CI_{m,n} \;=\; \int {\rd^D p\over (2\pi)^D}
         {(p^2-X)^m\over \left( 2m\varepsilon(p, \phi)\right)^n}, 
\label{cI-mn}
\end{equation}
so one can write 
\be
  \CV_1(\mu,\phi)={1\over 4m} \CI_{0,-1}. 
  \ee
 \begin{figure}[!ht]
\leavevmode
\begin{center}
\includegraphics[height=4cm]{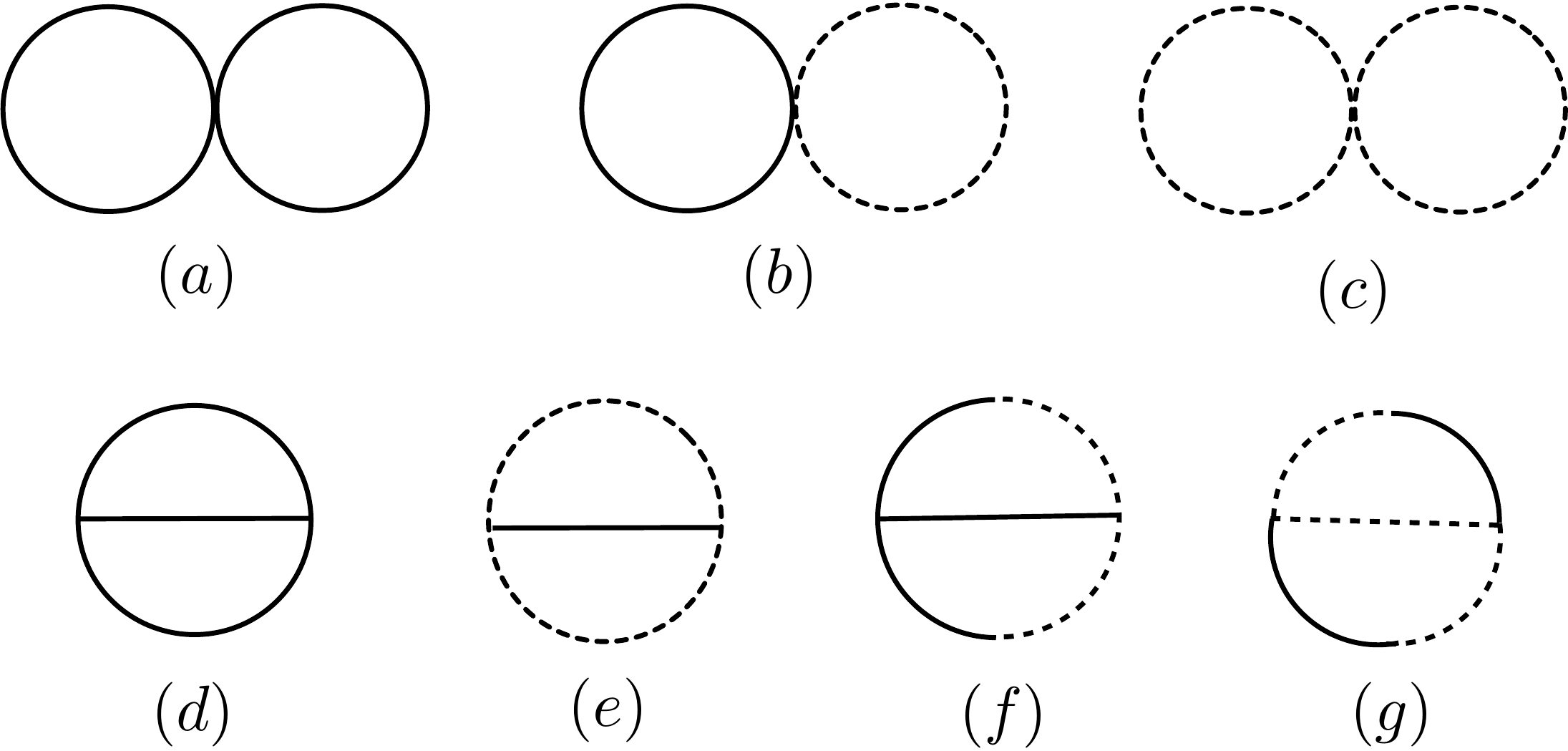}
\end{center}
\caption{Two-loop diagrams for the Lieb--Liniger model.}
\label{twoloop}
\end{figure} 
Finally, the two-loop potential $\CV_2(\mu,\phi)$ is the sum of the contributions of the vacuum diagrams shown in \figref{twoloop}. The diagrams 
in the first line involve products of one-loop integrals:
\be
      \CV_2^{(a)}  =  {3 \over 64} g\, \CI_{1,1}^2, \qquad  
\CV_2^{(b)}  =  {1 \over 32} g\, \CI_{-1,-1} \CI_{1,1}, \qquad 
\CV_2^{(c)}  =  {3 \over 64} g\, \CI_{-1,-1}^2.
\ee
The diagrams in the second line involve the integrals 
\be
\ba
\CJ_{l,m,n} 
= \int {\rd^D p\over (2\pi)^D}
         \int {\rd^D q\over (2\pi)^D} &
       \left( {p^2-X \over 2m \, \varepsilon(p, \phi)}\right)^l
        \left( {q^2-X \over 2m\, \varepsilon(q, \phi)}\right)^m
         \left( {r^2-X \over 2m\, \varepsilon(r, \phi)}\right)^n\\
         & \times 
        {1  \over 2m( \varepsilon(p,\phi)+\varepsilon(q,\phi)+\varepsilon(r,\phi) )},
        \ea
\label{CJ-lmn}
\ee
where $r=|{\bf p}+{\bf q}|$. They are given by 
\be
\ba
\CV_2^{(d)} & = - {3 \over 16} m g^2 \phi^2 \CJ_{1, 1, 1}, 
\qquad  \CV_2^{(e)} = - {1 \over 16} m g^2 \phi^2 \CJ_{-1, -1, 1},\\
\CV_2^{(f)} &=  {3 \over 8} m g^2 \phi^2 \CJ_{0, 0, 1},
 \qquad 
\CV_2^{(g)}  = - {1 \over 8} m g^2 \phi^2 \CJ_{-1,0, 0}. 
\ea
\ee

We can now find the minimum $\phi$ solving (\ref{Omega-min}), which we also expand according to the loop order as 
\be
\phi= \phi_\star + \phi_1+ \cdots
\ee
The classical vev $\phi_\star$ minimizes (\ref{ome0}) and one finds
\be
\label{v0}
 \phi_\star^2 = {2 \mu \over g}. 
 \ee
For this value of $\phi$, the parameter $X$ in (\ref{X-def}) vanishes. The field $\eta$ is massless and is the Goldstone boson of the model. The propagator simplifies 
when $X=0$ and one recovers Bogoliubov's
dispersion relation:
\begin{equation}
\varepsilon(k, \phi_\star )=
      {k\sqrt{k^2+\Lambda^2}\over 2m}, 
\label{Bogol}
\end{equation}
where 
\be
\Lambda^2=4m\mu.
\ee
However, it is easy to see that when $D=1$ and $X=0$ many of the integrals above are IR divergent. We will then proceed as we have done in the relativistic case 
and we will introduce an explicit IR regulator by setting 
\be
X= -\epsilon^2. 
\ee
We will take $\epsilon \to 0$ at the end of the calculation. Let us then introduce the regulated dispersion relation 
\begin{equation}
\varepsilon(k; \epsilon) =
        {{\sqrt{k^2 + \epsilon^2}} \sqrt{k^2+\Lambda^2}\over 2m}, 
        \label{reg-bogol}
\end{equation}
the regulated one-loop integrals
\be
I_{m,n}(\Lambda, \epsilon)= \int{\rd^D p\over (2\pi)^D}
         {(p^2+\epsilon^2)^m\over (2m\, \varepsilon(p; \epsilon))^n}, 
\label{rI-mn}
\end{equation}
and the regulated two-loop integrals
\be
\ba
J_{l,m,n} (\Lambda, \epsilon)
= \int{\rd^D p\over (2\pi)^D}
         \int{\rd^D q\over (2\pi)^D} &
       \left( {p^2+\epsilon^2 \over 2m \, \varepsilon(p; \epsilon)}\right)^l
        \left( {q^2+\epsilon^2 \over 2m\, \varepsilon(q; \epsilon)}\right)^m
         \left( {r^2+\epsilon^2 \over 2m\, \varepsilon(r; \epsilon)}\right)^n\\
         & \times 
        {1  \over 2m( \varepsilon(p; \epsilon)+\varepsilon(q; \epsilon)+\varepsilon(r; \epsilon) )}.
        \ea
\label{J-lmn}
\ee
We can now express all results in terms of these regulated integrals. 
The one-loop correction to the condensate is given by \cite{b-nieto}
\begin{equation}
\label{phi1-ll}
\phi_1=- {g \phi_\star \over 16 \mu} (3 I_{1,1} + I_{-1,-1}), 
\end{equation}
and the final result for the grand potential up to two-loops is as in \cite{b-nieto}, 
\begin{equation}\label{Omu}
\Omega(\mu) = -{\mu^2\over g}   
    + {1\over 4m}I_{0,-1}
    + {mg\mu\over 8} J
    + {g\over 32} \left(
      I_{-1,-1}^2 - 2 I_{-1,-1} I_{1,1} - 3 I_{1,1}^2 \right), 
\end{equation}
where
\begin{equation}
J(\epsilon) \;=\; 6 J_{0, 0, 1} - J_{-1, -1, 1} 
        - 3 J_{1, 1, 1}  - 2 J_{-1, 0, 0}, 
\label{J-def}
\end{equation}
and it is independent of $\Lambda$. Of course, all the integrals have to be understood as IR regularized integrals. 

Let us analyze these integrals in more detail. It turns out that, in $D=1$, $I_{0, -1}$ and $I_{1,1}$ are IR convergent but UV divergent. 
After dimensional regularization they lead to finite results that can be obtained explicitly from the formulae in \cite{b-nieto}. One finds, 
\be
\label{id}
I_{0,-1}(\Lambda, 0)= -{\Lambda^3\over 3 \pi}, \qquad I_{1,1}(\Lambda, 0)=-{\Lambda \over \pi}. 
\ee
On the other hand, $I_{-1,-1}(\Lambda, \epsilon)$ is both UV divergent and IR divergent as $\epsilon \rightarrow 0$. It can be computed 
for arbitrary $\epsilon$ in dimensional regularization, and one obtains a finite result for $D=1$ and $\epsilon>0$. Its expansion as 
$\epsilon \rightarrow 0$ can be calculated analytically and one finds, 
\be
\label{i3}
I_{-1,-1}(\Lambda, \epsilon)= \Lambda \left\{ -{1\over 2 \pi} \log\left( {\epsilon^2 \over \Lambda^2} \right) + {2 \log(2) -1 \over \pi} + \CO(\epsilon^2)\right\} . 
\ee
Note that $\phi_1$, given in (\ref{phi1-ll}), inherits this divergence, 
which has exactly the same functional form as in the relativistic case (\ref{phi1loop}). 
It is also easy to see that the two-loop integrals $J_{-1,-1, 1}$ and $J_{-1, 0,0}$ are both IR divergent. The integral $J_{-1,-1, 1}$ comes from the diagram (e) in 
\figref{twoloop}, which is the counterpart of the IR divergent diagram also labelled as (e) in \figref{f-2loops}. The integral $J_{-1,0,0}$ appears in the diagram (f) in 
\figref{twoloop}. It involves the ``mixed" propagator of the $\xi$ and $\eta$ fields, and it has no counterpart in the relativistic $O(N)$ theory. However, as in \cite{jevicki}, 
all divergences appearing in the calculation of the grand potential should cancel 
order by order in the $g$ expansion. Using the values of the integrals (\ref{id}) and (\ref{i3}), and setting $m=1/2$ for convenience, we find the 
following result for the grand potential at two loops: 
\be
\label{pt}
\ba
\Omega(\mu)&=-{\mu^2 \over g}-{\sqrt{2} \over 3 \pi} \mu^{3/2}\\
& + {\mu g \over 16} \left( J(\epsilon) + {1\over 4 \pi^2} \log^2(\epsilon^2) - {2 \log(2) \over \pi^2} \log(\epsilon^2)+ {4 \log^2(2) \over  \pi^2} 
-{4 \over \pi^2} \right)+\cdots 
\ea
\ee
We note that the integral $J(\epsilon)$ can be further simplified and put into the form 
\be
\ba
J(\epsilon)&={1\over 4 \pi^2} \int \rd p\, \rd q\, \rd r\, \delta(p+q+r) F(p,q,r;\epsilon)\\
&={1\over  \pi^2}\int_0^\infty \rd p\int_0^p \rd q \left(F(p,q,p+q;\epsilon)+F(p,q,p-q;\epsilon)\right), 
\ea
\ee
where
\be
F(p,q,r;\epsilon)= -{1\over 3} {1\over E_p+E_q+E_r}{1\over \alpha_p \alpha_q \alpha_r} \left(\alpha_p \alpha_q+ \alpha_p \alpha_r + \alpha_q \alpha_r-3 \right)^2, 
\ee
and
\be
E_k=\sqrt{k^2+1}\sqrt{k^2+\epsilon^2}, \qquad 
\alpha_k=\sqrt{\frac{k^2+1}{k^2+\epsilon^2}}. 
\ee
The result (\ref{pt}) agrees with (\ref{bethe}) up to one-loop. In order to have agreement up to two-loops, we must have
\be
\label{an-pred}
 \lim_{\epsilon \to 0} \left( J(\epsilon) + {1\over 4 \pi^2} \log^2(\epsilon^2) 
 - {2 \log(2) \over \pi^2} \log(\epsilon^2)  \right)={1\over 3}- {4 \log^2(2) \over \pi^2}.
 \ee
 We have not attempted to establish (\ref{an-pred}) analytically. Numerically, we find 
 \be
 \lim_{\epsilon \to 0} \left( J(\epsilon) + {1\over 4 \pi^2} \log^2(\epsilon^2) - {2 \log(2) \over \pi^2} \log(\epsilon^2)  \right)\approx 1.38613061...
 \ee
 which agrees with the r.h.s. of (\ref{an-pred}) in all stable numerical digits.  

In order to establish the existence of an IR renormalon, one should find a calculable family of diagrams 
which leads to an IR finite answer and which grow factorially with the number of loops. In the case of the relativistic $O(N)$ scalar theory, 
the choice of such a set of diagrams was done for us by the large $N$ expansion. There is a natural $U(N)$ generalization of the theory with action (\ref{actpsi}) 
(see e.g. \cite{andersen-largeN, nogueira, pastu-N}), but in $D=1$ the dominant and subdominant diagrams at large $N$ do not contribute to the ground state energy 
(as expected from \cite{yang}), and the $1/N$ expansion is of little help in this case. It would be very interesting to complete this study by finding a 
sequence of diagrams which leads to an IR renormalon and explains the large order behavior (\ref{om-as}).

\sectiono{Conclusions}
\label{sec-con}

In this paper we have studied super-renormalizable scalar field theories in two-dimensions with a spontaneously broken $O(N)$ symmetry at the 
classical level. As noted long ago by Jevicki, one can obtain an IR finite perturbative series around the classical 
vacuum of these theories. These series give the correct weak-coupling, asymptotic expansion of the observables. However, we have shown by an explicit large $N$ calculation that 
the resulting series for the ground state energy is not Borel summable and leads to an IR renormalon singularity. 

This IR renormalon qualifies the assertion often found in the literature according to which 
there are no renormalons in super-renormalizable theories. We have argued that the physical reason behind this renormalon is that, as a consequence of the Coleman--Mermin--Wagner theorem, 
one is expanding around a ``false vacuum". Although the IR divergences due to the would-be Goldstone bosons cancel, 
the perturbative series remains ``IR sensitive" \cite{beneke} and leads to the IR renormalon.

The $O(N)$ model can be described at low energies by the non-linear sigma model \cite{zjbook}. Since the latter 
has IR renormalons, this would explain the appearance of an IR renormalon 
in the former\footnote{We would like to thank Lorenzo di Pietro and Marco Serone 
for pointing this out.}. It would be illuminating to make this more precise, but it seems clear that 
the $O(N)$ model provides a much simpler realization of IR renormalons than 
the non-linear sigma model, precisely due to its simpler UV behavior. In particular, we do not expect the 
$O(N)$ model to have UV renormalons, while observables in the non-linear sigma model 
display a complicated mixing of both IR and UV renormalons (see e.g. \cite{mr-ren} for an example).

It is also interesting to compare the non-Borel summable perturbative series we find in the $O(N)$ model 
with the one around the perturbative vacua in the double-well potential. In quantum mechanics, the parity symmetry of this 
potential can not be spontaneously broken, and the perturbative vacua are false vacua. The lack of Borel summability of the resulting perturbative 
series reflects the wrong choice of vacuum that one has made to begin with. In this quantum-mechanical example, the factorial divergence is 
an instanton effect and can be cured by taking into account multi-instanton sectors (see e.g. \cite{zjj1, zjj2}). In the case of the $O(N)$ theory, we are also expanding around a false vacuum, 
due to the Coleman--Mermin--Wagner theorem. The lack of Borel summability might be interpreted as the price to pay for committing 
the original sin of expanding around this incorrect vacuum.  
However, as we showed in this paper, the resulting Borel singularity is in this case an IR renormalon effect. 

Our finding further supports the idea that renormalons should not be associated exclusively to 
renormalizable theories. Renormalon singularities have been recently found in different contexts, like quantum mechanics \cite{rqm} 
and various condensed matter models \cite{mr-long}. In the case studied in this paper, as well as in the examples of \cite{mr-long}, 
renormalon singularities are due to the integration over momenta 
in ``dangerous" regions, which are not even necessarily the IR or UV regions (for example, in many-fermion models, 
these regions are often associated to the Fermi surface). 

The IR renormalon unveiled in this paper is perhaps one of the simplest incarnations of a 
renormalon singularity. It shows that the perturbative approach to the two-dimensional 
$O(N)$ quantum field theory is insufficient, and one has to take into 
account some sort of non-perturbative effect in order to make sense of perturbation theory. Another manifestation 
of the simplicity of this model is that it is possible to make a very detailed analysis of its resurgent structure. In particular, 
the associated trans-series can be fully determined, and we have presented a conjectural form for the exact Borel transform 
which leads to an explicit expression for the original perturbative series in terms of even zeta functions. However, it is not clear to us what is the nature of the  
non-perturbative sectors that lead to the exponentially small corrections appearing in the trans-series. Finding an explicit 
description for them is probably the most important problem opened by this investigation. 

There are other open problems that should be addressed. First of all, it would be very 
interesting to (dis)prove our conjecture about the existence 
of a similar renormalon in the Lieb--Liniger model. Note that in this case, with the help of the exact Bethe 
ansatz solution, one could try to construct the appropriate trans-series for the ground state energy. 
It would be also interesting to find other examples of super-renormalizable field theories with a 
Coleman--Mermin--Wagner ``false vacuum" where 
one can study similar renormalon singularities in various observables. 
Since our result is based on a large $N$ analysis, it would be very useful 
to find an example where the large order behavior of the perturbative series 
at finite $N$ can be studied explicitly and it is controlled by a renormalon of this type. 

Another natural line of enquiry, in view of recent work \cite{tin,circle,yayo}, is to understand the fate of the IR renormalon found 
in this paper after a (twisted) compactification on a circle. Perhaps the approach of 
\cite{a-unsal-long, dunne-unsal-cpn,cherman-dorigoni-dunne-unsal, cdu, misumi-1,du-on, misumi-2} can be also 
applied in this case, and provide a semiclassical interpretation of this renormalon in a suitable compactification of the theory. It might also 
happen that compactification makes the renormalon disappear completely, as argued in \cite{tin}. In fact, as we mentioned above, the $O(N)$ theory 
on an AdS$_2$ space of sufficiently small radius will have a vacuum where the symmetry is spontaneously broken. If our picture is correct, in that case we expect the 
perturbative series to be Borel summable again, with no trace of IR renormalons left. An explicit test of this expectation would be of great interest. 

\section*{Acknowledgements}
We would like to thank Jens Andersen, Volodymyr Pastukhov, Santi Peris, Lorenzo di Pietro, Marco Serone and Peter Wittwer for 
useful conversations and correspondence. We are particularly grateful to Gerald Dunne, 
Antal Jevicki, Santi Peris and Marco Serone for their 
comments on a preliminary version of this paper. This work has been supported in part by the Fonds National Suisse, 
subsidy 200021-175539, by the NCCR 51NF40-141869 ``The Mathematics of Physics'' (SwissMAP), 
and by the ERC Synergy Grant ``ReNewQuantum".

\bibliographystyle{JHEP}

\linespread{0.6}
\bibliography{biblio-2drenormalons}

\end{document}